\documentclass[english,aps,pra,showpacs,twocolumn]{revtex4-1}
\usepackage[utf8]{inputenc}
\usepackage{color}
\usepackage{babel}

\usepackage{float}
\usepackage{amsmath}
\usepackage{graphicx}
\usepackage{amssymb}
\usepackage[unicode=true, pdfusetitle,
 bookmarks=false,
 breaklinks=true,pdfborder={0 0 1},backref=false,colorlinks=true]
 {hyperref}

\makeatletter

\providecommand{\tabularnewline}{\\}

\@ifundefined{textcolor}{}
{%
 \definecolor{BLACK}{gray}{0}
 \definecolor{WHITE}{gray}{1}
 \definecolor{RED}{rgb}{1,0,0}
 \definecolor{GREEN}{rgb}{0,1,0}
 \definecolor{BLUE}{rgb}{0,0,1}
 \definecolor{CYAN}{cmyk}{1,0,0,0}
 \definecolor{MAGENTA}{cmyk}{0,1,0,0}
 \definecolor{YELLOW}{cmyk}{0,0,1,0}
 }

\usepackage{slashbox}
\RequirePackage{mathrsfs}
\usepackage{bm}
\usepackage{txfonts}
\usepackage{dcolumn}

\newcommand{\llll}{\raisebox{-0.4\height}{\includegraphics[height=1cm]{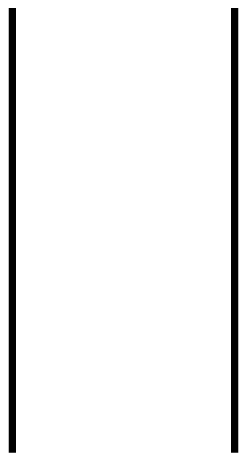}}}
\newcommand{\lcrosslU}{\raisebox{-0.4\height}{\includegraphics[height=1cm]{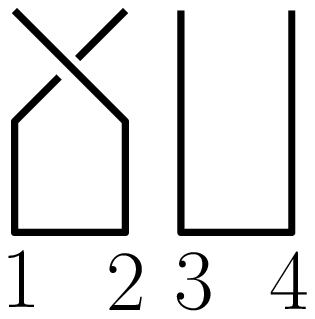}}}
\newcommand{\complex}{\raisebox{-0.4\height}{\includegraphics[height=2cm]{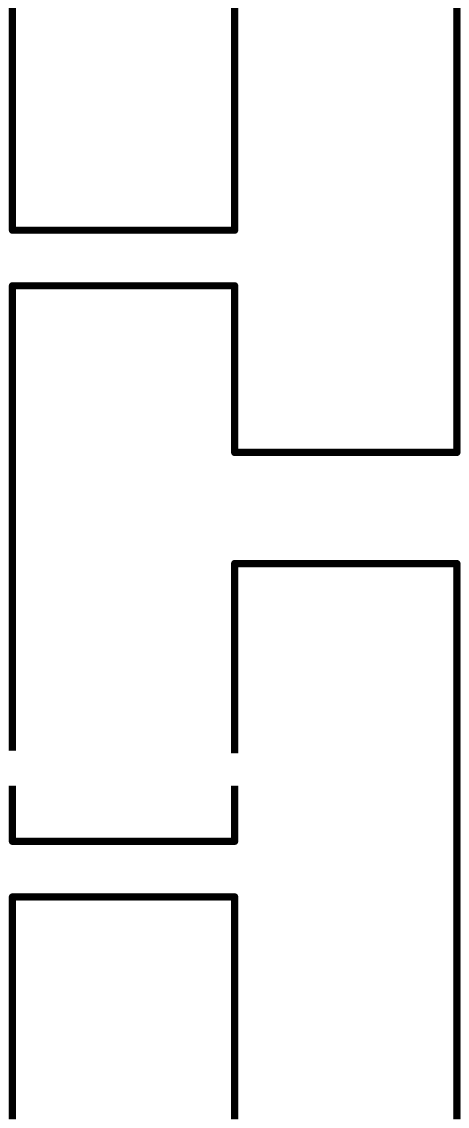}}}
\newcommand{\cross}{\raisebox{-0.4\height}{\includegraphics[height=1cm]{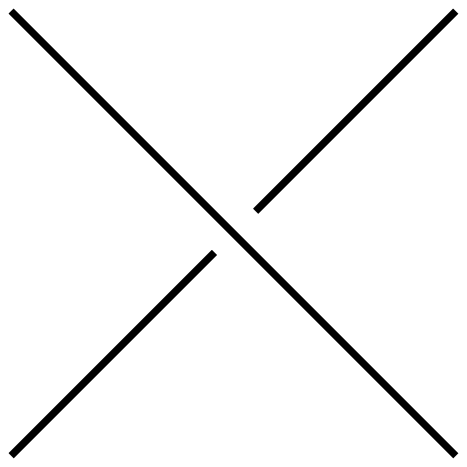}}}
\newcommand{\myloop}{\raisebox{-0.33\height}{\includegraphics[height=0.5cm]{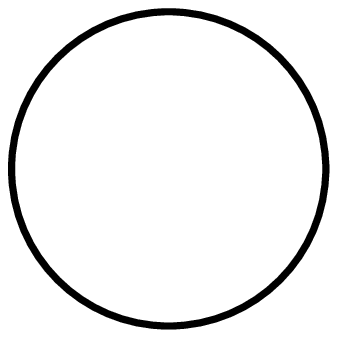}}}
\newcommand{\oUn}{\raisebox{-0.4\height}{\includegraphics[height=1cm]{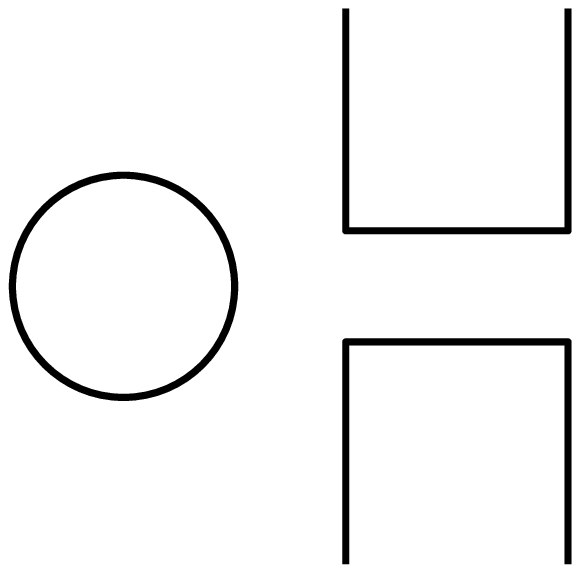}}}
\newcommand{\twocross}{\raisebox{-0.4\height}{\includegraphics[height=1cm]{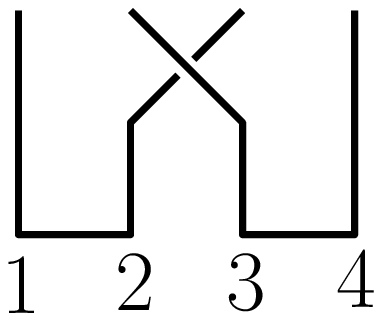}}}
\newcommand{\twojoin}{\raisebox{-0.4\height}{\includegraphics[height=1cm]{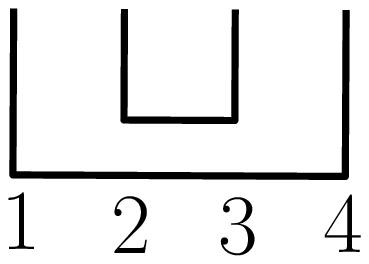}}}
\newcommand{\twosep}{\raisebox{-0.4\height}{\includegraphics[height=1cm]{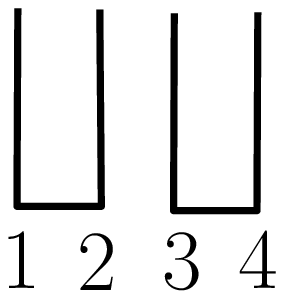}}}
\newcommand{\Un}{\raisebox{-0.4\height}{\includegraphics[height=1cm]{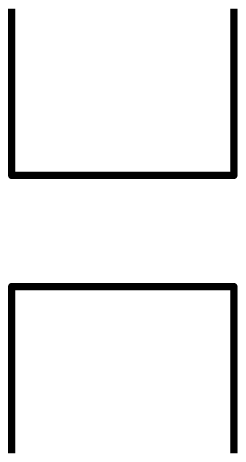}}}
\newcommand{\Unl}{\raisebox{-0.4\height}{\includegraphics[height=1cm]{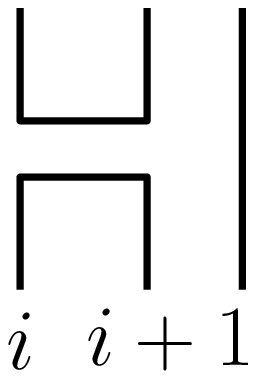}}}

\newcommand{\Uon}{\raisebox{-0.4\height}{\includegraphics[height=1cm]{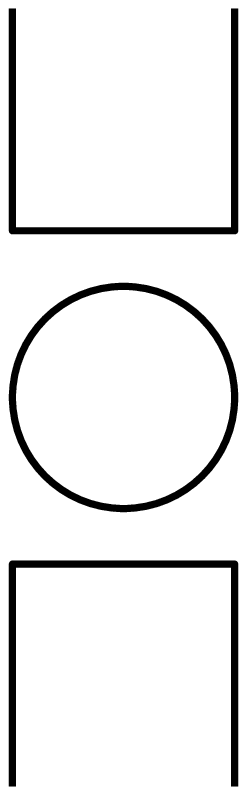}}}
\newcommand{\nn}{\raisebox{-0.4\height}{\includegraphics[height=1cm]{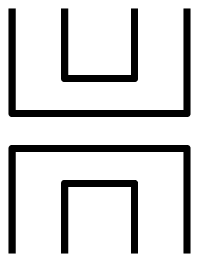}}}
\newcommand{\nm}{\raisebox{-0.4\height}{\includegraphics[height=1cm]{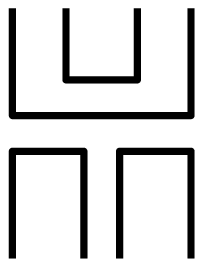}}}
\newcommand{\mn}{\raisebox{-0.4\height}{\includegraphics[height=1cm]{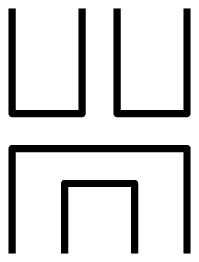}}}
\newcommand{\mm}{\raisebox{-0.4\height}{\includegraphics[height=1cm]{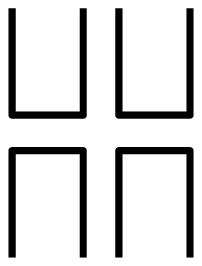}}}
\newcommand{\Unabcd}{\raisebox{-0.4\height}{\includegraphics[height=1cm]{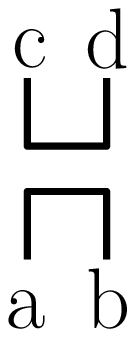}}}
\newcommand{\twoS}{\raisebox{-0.4\height}{\includegraphics[height=1cm]{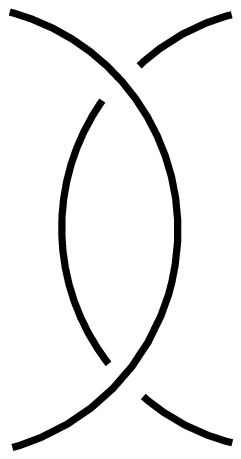}}}
\newcommand{\twoSS}{\raisebox{-0.4\height}{\includegraphics[height=1cm]{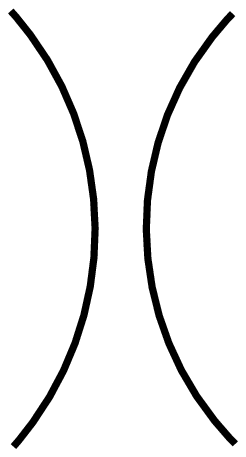}}}
\newcommand{\Unij}{\raisebox{-0.6\height}{\includegraphics[height=1cm]{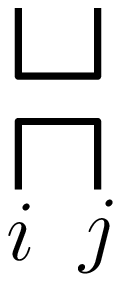}}}
\newcommand{\uij}{\raisebox{-0.3\height}{\includegraphics[height=0.6cm]{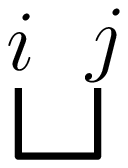}}}
\newcommand{\nij}{\raisebox{-0.3\height}{\includegraphics[height=0.6cm]{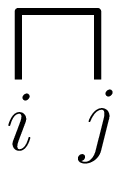}}}

\makeatother

\begin{document}


\title{The $\ell_{1}$-norm in quantum information via the approach of Yang-Baxter
Equation}

\author{Kai Niu$^{1}$}

\email{niukai@nankai.edu.cn}

\author{Kang Xue$^{2}$}

\author{Qing Zhao$^{3}$}

\author{Mo-Lin Ge$^{1}$}

\email{geml@nankai.edu.cn}

\affiliation{$^{1}$ Theoretical Physics Section, Chern Institute of Mathematics,
Nankai University, Tianjin, 300071, China\\
$^{2}$ Department of Physics, Northeast Normal University, Changchun,
Ji Lin, 120024, China\\
$^{3}$ Department of Physics, College of Science, Beijing Institute
of Technology, Beijing, 100081, China}

\date{\today}
\begin{abstract}
The role of $\ell_{1}$-norm in Quantum Mechanics (QM) has been
studied through Wigner's D-functions where $\ell_{1}$-norm means
$\sum_{i}\left|C_{i}\right|$ for $\left|\Psi\right\rangle
=\sum_{i}C_{i}\left|\psi_{i}\right\rangle $ if
$\left|\psi_{i}\right\rangle $ are uni-orthogonal and normalized
basis. It was shown that the present two types of transformation
matrix acting on the natural basis in physics consist in an unified
braiding matrix, which can be viewed as a particular solution of the
Yang-Baxter equation (YBE). The maximum of the $\ell_{1}$-norm is
connected with the maximally entangled states and topological
quantum field theory (TQFT) with two-component anyons while the
minimum leads to the permutation for fermions or bosons.
\end{abstract}


\maketitle

\section{Introduction: two types of braiding matrices, Yang-Baxter equation
and Temperley-Lieb algebra\label{sec:Introduction}}

The purpose of this paper is committed to clarifying how
$\ell_{1}$-norm participates in Quantum Mechanics (QM) and
demonstrating the physical meaning through acceptable physical
examples. In QM, a wave function $\left|\Psi\right\rangle $ can be
decomposed to $\left|\Psi\right\rangle
=\sum_{i}C_{i}\left|\psi_{i}\right\rangle $, where
$\left|\psi_{i}\right\rangle $ is uni-orthogonal basis and the
normalizability of $\left|\Psi\right\rangle $ reads
\begin{align} \langle\Psi\left|\Psi\right\rangle  &
=\sum_{i}\left|C_{i}\right|^{2}=1.\end{align} We call
$\sum_{i}\left|C_{i}\right|^{2}=\left\Vert C\right\Vert _{\ell_{2}}$
as $\ell_{2}$-norm, which indicates the square integrability of the
wave function. Meanwhile the notation
$\sum_{i}\left|C_{i}\right|=\left\Vert C\right\Vert _{\ell_{1}}$ is
called $\ell_{1}$-norm.

We may ask whether an $\ell_{1}$-norm $f=\sum_{i}\left|C_{i}\right|$
plays role in QM and if so, which physical model represents this
statement. For this target, we should go a long way. We shall show
that the local maximum and minimum of $\ell_{1}$-norm will lead to
two types of braiding matrices that have existed in physics. One is
related to the entangled states including the anyonic description
{\cite{1,2,3}}, and the other to the permutation type {\cite{4}},
which lays down the base of solvable models exactly {\cite{4,5}}. In
order to explain the matter clearly, we have to begin with the braid
relation, Yang-Baxter Equation (YBE), and their particular matrix
forms. And then the physical consequence of extremism of
$\ell_{1}$-norm was explained.

Recently, a new development has been used to connect the braid
matrix, as well as the YBE, with the entangled states
{\cite{6,7,8,9,10,11}. We start the discussion with the maximally
entangled states, i.e., the Bell states. For a two-qubit system,
Bell states are defined by:\begin{align*}
\left|\Phi^{\pm}\right\rangle  & =\frac{1}{\sqrt{2}}\left(\left|\uparrow\uparrow\right\rangle \pm\left|\downarrow\downarrow\right\rangle \right),\\
\left|\Psi^{\pm}\right\rangle  &
=\frac{1}{\sqrt{2}}\left(\left|\uparrow\downarrow\right\rangle
\pm\left|\downarrow\uparrow\right\rangle \right).\end{align*} The
Bell states are connected to the natural basis
$\left|\psi_{0}\right\rangle =(\left|\uparrow\uparrow\right\rangle
,\left|\uparrow\downarrow\right\rangle
,\left|\downarrow\uparrow\right\rangle
,\left|\downarrow\downarrow\right\rangle )^{T}$ by a unitary
transformation matrix $W$, which satisfies\begin{align}
\left(\Phi^{+},\Psi^{+},-\Psi^{-},-\Phi^{-}\right) &
=W\left(\left|\uparrow\uparrow\right\rangle
,\left|\uparrow\downarrow\right\rangle
,\left|\downarrow\uparrow\right\rangle
,\left|\downarrow\downarrow\right\rangle
\right)^{T},\label{eq:w}\end{align} where\begin{align*} W &
=\frac{1}{\sqrt{2}}\left(\begin{array}{cccc}
1 & 0 & 0 & 1\\
0 & 1 & 1 & 0\\
0 & -1 & 1 & 0\\
-1 & 0 & 0 & 1\end{array}\right).\end{align*} The $W$ can be
extended to matrix $b$ such as \cite{7,10}\begin{align}
b_{\mbox{I}}(q) & =\frac{1}{\sqrt{2}}\left(\begin{array}{cccc}
1 & 0 & 0 & q\\
0 & 1 & \epsilon & 0\\
0 & -\epsilon & 1 & 0\\
-q^{-1} & 0 & 0 & 1\end{array}\right)=\frac{1}{\sqrt{2}}(1+M),\label{eq:b_tilt}\\
\epsilon^{2} & =1,\quad M^{2}=-1,\quad q=e^{i\alpha}.\end{align}
Kauffman \emph{et al. }{\cite{7}} have shown that the matrix $W$ is
nothing but a braid matrix ($N^{2}=4$), which satisfies\begin{align}
B_{1}B_{2}B_{1} & =B_{2}B_{1}B_{2},\label{eq:bbb}\end{align}
where\begin{align*}
B_{1}^{\mbox{I}} & \equiv B_{12}^{\mbox{I}}=b_{\mbox{I}}(q)\otimes I,\\
B_{2}^{\mbox{I}} & \equiv B_{23}^{\mbox{I}}=I\otimes b_{\mbox{I}}(q).\end{align*}

On the other hand, in solving a one-dimensional (1D) model with
$\delta$-function potential {\cite{12}}, and a low-dimensional
statistical model, as well as the chain models, the other types of
braiding matrices were introduced years ago{\cite{13}}. The simplest
form is given by {\cite{14}}\begin{align}
b_{\mbox{II}}=\left(\begin{array}{cccc}
q & 0 & 0 & 0\\
0 & 0 & -\eta & 0\\
0 & -\eta^{-1} & q-q^{-1} & 0\\
0 & 0 & 0 & q\end{array}\right),\label{eq:type_2_b}\end{align} that
was known as the $q$-deformation of permutation. Here
$\eta=e^{i\alpha}$ with $\alpha$ being any flux, when $\eta=-1$,
$q=1$, Eq. (\ref{eq:type_2_b}) reduces to the permutation, which is
universal symmetry operator for identical particles either boson or
fermion. A braiding matrix can be viewed as the asymptotic behavior
of $2$-body scattering matrix, i.e., the momenta independent part of
S-matrix. For a given matrix satisfying Eq. (\ref{eq:bbb}), the
corresponding $\breve{R}(x)$-matrix can obey\begin{align}
\breve{R}_{1}(x)\breve{R}_{2}(xy)\breve{R}_{1}(y) &
=\breve{R}_{2}(y)\breve{R}_{1}(xy)\breve{R}_{2}(x),\label{eq:YBE_org}\end{align}
where $x$ is spectral parameter related to 1D momenta($u$ for
$x=e^{iu}$) which obeys the conservation law, and \begin{align*}
\breve{R}_{1}(x)\equiv\breve{R}_{12}(x) & =\breve{R}(x)\otimes I,\\
\breve{R}_{2}(x)\equiv\breve{R}_{23}(x) &
=I\otimes\breve{R}(x).\end{align*} Obviously, matrix $B$ is a
particular case of $\breve{R}(x)$. The physical meaning of
$\breve{R}(x)$ is the S-matrix of $2$-body scattering. Eq.
(\ref{eq:YBE_org}) means that if any $3$-body scattering can be
decomposed to three $2$-body ones, then two collision ways should be
equal to each other. For a given $B$ to find $\breve{R}(x)$ is
called Yang-Baxterization {\cite{4,14}}. It is easy to be made if
$B$ (hence $\breve{R}(x)$) does have two distinct eigenvalues.

The Yang-Baxter Equation (YBE) originally was introduced to solve
the one-dimensional $\delta$-interaction models {\cite{12}}, and the
statistical models on lattices {\cite{13}}. The importance of the
YBE is further revealed as a beginning for the method of quantum
inverse scattering {\cite{4,14}}. YBE also plays an important role
in solving the integrable models in quantum field theory and exactly
solvable models in statistical mechanics. In quantum field theory,
the YBE is used to describe the scattering of particles in ($1+1$)
dimensions. The basic concept of the YBE is to factorize the
three-body scattering into two-body scattering processes. The YBE is
also very useful in completing integrable statistical models, whose
solutions can be found by means of the nested Bethe ansatz
{\cite{15}}.

Observing the two different types of braiding matrices
$b_{\mbox{I}}$ and $b_{\mbox{II}}$, both of them can be expressed in
terms of matrix $T$ such as \begin{align} S &
=\rho(1+fT).\label{eq:S_org}\end{align} where $S$ can be either
$b_{\mbox{I}}$ or $b_{\mbox{II}}$. Constant $f$ and matrix $T$ can
be defined through (\ref{eq:b_tilt}) or (\ref{eq:type_2_b}). For
type I, we have\begin{align} T_{\mbox{I}} &
=\frac{1}{\sqrt{2}}\left(\begin{array}{cccc}
1 & 0 & 0 & e^{i\alpha}\\
0 & 1 & -i\epsilon & 0\\
0 & i\epsilon & 1 & 0\\
e^{-i\alpha} & 0 & 0 &
1\end{array}\right),\quad(\epsilon^{2}=1),\label{eq:type_1_T}\end{align}
and for type II\begin{align} T_{\mbox{II}}^{'} &
=\left(\begin{array}{cccc}
0 & 0 & 0 & 0\\
0 & 1 & \eta & 0\\
0 & \eta^{-1} & 1 & 0\\
0 & 0 & 0 & 0\end{array}\right),\label{eq:type-2-T_prime}\end{align}
\begin{align}
\mbox{or } & T_{\mbox{II}}=\left(\begin{array}{cccc}
1 & 0 & 0 & \eta\\
0 & 0 & 0 & 0\\
0 & 0 & 0 & 0\\
\eta^{-1} & 0 & 0 & 1\end{array}\right),\quad V=\left(\begin{array}{cccc}
0 & 1 & 0 & 0\\
1 & 0 & 0 & 0\\
0 & 0 & 0 & 1\\
0 & 0 & 1 & 0\end{array}\right).\label{eq:type-2-T}\end{align}
Noting that $T_{\mbox{II}}=VT_{\mbox{II}}^{'}V^{\dagger}$ is still
the solution of Eq. (\ref{eq:bbb}) through Eq. (\ref{eq:S_org}).
Both $T_{\mbox{I}}$ and $T_{\mbox{II}}$, and their extensions have
nice properties, i.e., they satisfy the relations\begin{align}
T_{i}^{2}=dT_{i} & ,\quad(T_{i}\equiv T_{i\ i+1})\label{eq:TL-1}\\
T_{i}T_{i+1}T_{i} & =T_{i}.\label{eq:TL-2}\end{align} where $d$ is
constant. The relations which satisfy Eqs. (\ref{eq:TL-1}) and
(\ref{eq:TL-2}) is called Temperley-Lieb algebra (T-L) {\cite{16}}
that originated in spin chain model. The $T_{i}$ can be operators to
act on any dimensional models. The $T_{\mbox{I}}$ and
$T_{\mbox{II}}$ given by Eqs. (\ref{eq:type_1_T}) and
(\ref{eq:type-2-T}) are 4D representations of operator
$\hat{T_{i}}$.

There is a graphic expression of $\hat{T_{i}}$:\begin{align}
\hat{T}_{i} & \equiv\hat{T}_{i,i+1}=\Un,\nonumber \\
\hat{T}_{i}^{2} & =d\hat{T}_{i}=\Uon=\oUn\ ,\quad d=\myloop\ \mbox{(loop),}\label{eq:TL}\\
\hat{T}_{i} & \hat{T}_{i+1}\hat{T}_{i}=\complex=\Unl=T_{i}.\nonumber \end{align}

The matrix elements of operator $\hat{T}$ is
$(\hat{T}_{i})_{ab,cd}=\Unabcd$. With the operator $\hat{T}$, we
introduce the operator $\hat{S}(x)$, whose elements are formed by
matrix $\breve{R}(x)$:\begin{align} \hat{S}(x) &
=\rho\left[I+G(x)\hat{T}\right].\label{eq:S_YBE}\end{align} For
examples, the value of loop for $T_{\mbox{I}}$, $d=\sqrt{2}$,
whereas for $T_{\mbox{II}}$, $d=(q+q^{-1})$, i.e. $d=2$ at $q=1$. In
terms of Eq. (\ref{eq:TL}), the braiding matrix can be written as
the operator form:\begin{align} \hat{S} & =\rho(1+f\
\Un)=\cross.\label{eq:S_braid}\end{align} whose 4D matrix form is
given by Eqs. (\ref{eq:type_1_T}) and (\ref{eq:type-2-T}). In
(\ref{eq:S_braid}) a braiding means entangling. $\hat{S}$ means the
asymptotic behavior of S-matrix operator shown by over-crossing. The
under-crossing means $\hat{S}^{-1}$, i.e., $\twoS=\twoSS=I$. For
types I and II, there are only two distinct eigenvalues. Following
Kauffman {\cite{17}}, they have the decomposition:\begin{align}
\breve{S}= & \cross=\alpha\ \llll+\alpha^{-1}\Un,\end{align} It is
easy to find \begin{align*}
d=-\left(\alpha^{2}+\alpha^{-2}\right),\end{align*} and
then\begin{align*} f &
=\frac{1}{2}\left(-d\pm\sqrt{d^{2}-4}\right).\end{align*}
 in Eq. (\ref{eq:S_braid}). For type I ($d=\sqrt{2}$), $f_{\mbox{I}}=(-1)e^{\mp i\pi/4}$
while $f_{\mbox{II}}=-1$ at $q=1$ for type II ($d=2$).

Now we have expressed $2$-body scattering operator $\hat{S}(x)$
through operator $\hat{T}$ satisfying T-L algebra. It turns out in
Eq. (\ref{eq:S_YBE}) that operator $\hat{T}$ is nothing but the
scattering part in variable separation way. Eq. (\ref{eq:S_YBE})
describes a limited class of 1D scattering including a lot of exactly
solvable models connected with type II.

It is easy to establish the connection between the graphic
description and the spin operator. For instance, the operators
$T_{ij}$ for $T_{\mbox{I}}$ and $T_{\mbox{II}}$ take the
form:\begin{align*}
\hat{T}_{ij}^{\mbox{I}} & =\frac{1}{\sqrt{2}}\Big[I_{ij}+e^{i\alpha}S_{i}^{+}S_{j}^{+}+e^{-i\alpha}S_{i}^{-}S_{j}^{-}\\
 & \qquad+i\epsilon\left(S_{i}^{+}S_{j}^{-}-S_{i}^{-}S_{j}^{+}\right)\Big],\end{align*}
\begin{align*}
\hat{T}_{ij}^{\mbox{II}} & =\frac{1}{2}\left(I_{ij}+4S_{i}^{z}S_{j}^{z}\right)\\
 & \quad+e^{i\alpha}S_{i}^{+}S_{j}^{+}+e^{-i\alpha}S_{i}^{-}S_{j}^{-},\end{align*}
where $i$ and $j$ indicate the specified spaces and $\hat{T}_{ij}\left|k\right\rangle =\left|k\right\rangle $
for $k\neq i$, $k\neq j$.

By taking the elements $\left\langle
\psi_{0}\left|\hat{T}_{12}^{\mbox{I}}\right|\psi_{0}\right\rangle $
and $\left\langle
\psi_{0}\left|\hat{T}_{12}^{\mbox{II}}\right|\psi_{0}\right\rangle
$, we rederive Eqs. (\ref{eq:type_1_T}) and (\ref{eq:type-2-T}),
respectively.

The corresponding S-matrix (\ref{eq:S_YBE}) satisfies YBE for type I
of braiding matrices (\ref{eq:type_1_T}) given by
{\cite{10}}\begin{align*} \breve{R}_{\mbox{I}}(\theta,\alpha) &
=\left(\begin{array}{cccc}
\cos\theta & 0 & 0 & e^{i\alpha}\sin\theta\\
0 & \cos\theta & \sin\theta & 0\\
0 & -\sin\theta & \cos\theta & 0\\
-e^{-i\alpha}\sin\theta & 0 & 0 &
\cos\theta\end{array}\right),\end{align*} where
$\cos\theta=(1-x)/\sqrt{2(1+x^{2})}$ for type I. For type II
$x=e^{iu}$, the YBE is written in the form \begin{align*}
\breve{R}_{1}(u_{1})\breve{R}_{2}(u_{1}+u_{3})\breve{R}_{1}(u_{3}) &
=\breve{R}_{2}(u_{3})\breve{R}_{1}(u_{1}+u_{3})\breve{R}_{2}(u_{1}),\end{align*}
hence\begin{align*} \hat{S}(u) &
=\rho(u)\left[I+G(u)\hat{T}\right],\end{align*} the 4D
representation is
\begin{align*} \breve{R}^{\mbox{II}}=I+uP, & \quad
P(\eta)=\left(\begin{array}{cccc}
1 & 0 & 0 & 0\\
0 & 0 & \eta & 0\\
0 & -\eta^{-1} & 0 & 0\\
0 & 0 & 0 & 1\end{array}\right),\end{align*}
when $\eta=-1$, $P(\eta)$ is the $4\times4$ representation of permutation.

In short summary, besides the familiar
$\breve{R}^{\mbox{II}}(u)$-matrix related to chain models, we have
pointed out that the braiding matrix related to quantum information
has also its extension to satisfy YBE. Both of the two types of
braiding matrices obey the T-L algebra, which can be expressed in
terms of the graphic interpretation.

\section{\label{sec:Two-dimensional-braiding-matrices}Two-dimensional braiding
matrices and YBE}

In the above section, there are two types of 4D representations of
braiding matrices, hence the $\breve{R}(x)$-matrix has been shown.
In this section we shall review some results of 2D braiding
matrices, which obey the braid relation\begin{align*} ABA &
=BAB.\end{align*}

In order to keep the paper self-contained, we first explain the
basic concepts related to YBE. The Yang-Baxter matrix $R$ is a
$N^{2}\times N^{2}$ matrix acted on the tensor product space
$V\otimes V$, where $N$ is the dimension of $V$. Such a matrix $R$
satisfies the YBE:\begin{align}
R_{12}(u_{1})R_{23}(u_{2})R_{12}(u_{3}) &
=R_{23}(u_{3})R_{12}(u_{2})R_{23}(u_{1})\label{eq:YBE}\end{align}
where $R_{12}=R\otimes1$, $R_{23}=1\otimes R$, $u_{1},u_{2},u_{3}$
are spectral parameters. It should be noted that in the YBE of Eq.
(\ref{eq:YBE}), the spectral parameters are usually considered to be
related to the momenta and they must satisfy the conservation law,
i.e., $u_{2}$ is the addition of $u_{1}$ and $u_{3}$ either in
Lorentz form {\cite{10}} or in Galileo form, which depends on type I
or type II. When the parameters in the YBE take special value, the
Eq. (\ref{eq:YBE}) will reduce to the braid relation:\begin{align}
b_{12}b_{23}b_{12} & =b_{23}b_{12}b_{23},\label{eq:braid}\end{align}
where $b_{12}=b\otimes1$, $b_{23}=1\otimes b$ play similar action to
the matrices $R_{12}$ and $R_{23}$, but there is no parameter
dependence in this equation. In fact the braid relation
(\ref{eq:braid}) is the asymptotic form of the YBE (\ref{eq:YBE}).
It is also well known that such a braid relation can be reduced to a
$N\times N$ dimensional braid relation\begin{align} ABA &
=BAB.\label{eq:ABA_BAB}\end{align} A known example comes from the
conformal field theory (CFT) which is the simplification of
Nayak-Wilczek derivation of braiding matrices for fractional quantum
Hall effect (FQHE) {\cite{18,19}}.\begin{align*}
F_{I} & =\left[\frac{1}{w_{12}w_{34}(1-\xi)}\right]^{1/8}\left(1+\sqrt{1-\xi}\right)^{1/2},\\
F_{\Psi} & =\left[\frac{1}{w_{12}w_{34}(1-\xi)}\right]^{1/8}\left(1-\sqrt{1-\xi}\right)^{1/2},\\
\xi & =\frac{w_{12}w_{34}}{w_{13}w_{24}},\quad
w_{ij}=w_{i}-w_{j}.\end{align*} By setting $w_{1}=0$, $w_{2}=z$,
$w_{3}=1$, $w_{4}=w(\to\infty)$, we get $\xi=z(w-1)/(w-z)$ and
$\xi|_{w\to\infty}=z$. If we interchange the first two points
$w_{1}$, and $w_{2}$ (or $w_{3}$ and $w_{4}$), functions $F_{I}$ and
$F_{\Psi}$ will change to the superposition themselves. Through
calculations, it holds{\cite{18,19}}
\begin{align} \left.\left(\begin{array}{c}
F_{I}\\
F_{\Psi}\end{array}\right)\right|_{1\leftrightarrow2} & =e^{-i\pi/8}\left(\begin{array}{cc}
1 & 0\\
0 & i\end{array}\right)\left(\begin{array}{c}
F_{I}\\
F_{\Psi}\end{array}\right)=A\left(\begin{array}{c}
F_{I}\\
F_{\Psi}\end{array}\right),\label{eq:12}\\
\left.\left(\begin{array}{c}
F_{I}\\
F_{\Psi}\end{array}\right)\right|_{3\leftrightarrow4} & =\frac{e^{-i\pi/8}}{\sqrt{2}}\left(\begin{array}{cc}
1 & -i\\
-i & 1\end{array}\right)\left(\begin{array}{c}
F_{I}\\
F_{\Psi}\end{array}\right)=B\left(\begin{array}{c}
F_{I}\\
F_{\Psi}\end{array}\right).\label{eq:34}\end{align} The matrixes $A$
and $B$ are found to suit Eq. (\ref{eq:ABA_BAB}).

More generally the picture shown by $F_{I}$ and $F_{\Psi}$ can be
extended to the topological basis {\cite{3,6}} through the graphs,
if the T-L algebra is satisfied. For instance, the basis can be
introduced: \begin{align}
\left|e_{1}\right\rangle  & =\frac{1}{d}\twosep,\\
\left|e_{2}\right\rangle  &
=\frac{\epsilon}{\sqrt{d^{2}-1}}\left(\twojoin-\frac{1}{d}\twosep\right),\end{align}
where $\epsilon=\pm1$, $\left|e_{1}\right\rangle $ and
$\left|e_{2}\right\rangle $ are uni-orthonormalized basis. By making
the braiding between 1 and 2, 2 and 3, it forms the simplest
topological quantum field theory (TQFT). We introduce the braiding
operations $\hat{A}$ and $\hat{B}$, such as

\begin{align}
\hat{A}: & \quad\lcrosslU\quad\mbox{braiding the particles \mbox{1} and \mbox{2}},\label{eq:operator_A}\\
\hat{B}: & \quad\twocross\quad\mbox{braiding the particles \mbox{2}
and \mbox{3}}.\label{eq:operator_B}\end{align} The braiding cross
$\cross$ in $\hat{A}$ and $\hat{B}$ can be decomposed as {\cite{17}}
\begin{align} \breve{R}= & \cross=\alpha\
\llll+\alpha^{-1}\Un,\qquad
d=-\left(\alpha^{2}+\alpha^{-2}\right).\label{eq:R_decomposition}\end{align}
It is worth paying attention that $\left|e_{1}\right\rangle $ and
$\left|e_{2}\right\rangle $ occupy four spaces. Each crossing given
by Eq. (\ref{eq:R_decomposition}) means $4\times4$ representation of
braiding matrix . For both of the types we act the operator
$\hat{T}$ of Eq. (\ref{eq:TL}) on $\left|e_{1}\right\rangle $ and
$\left|e_{2}\right\rangle $, which lead to the two-dimensional
representations of $\hat{T}$:
\begin{align}
\hat{T}_{12}|e_{1}\rangle=\hat{T}_{34}|e_{1}\rangle & =d|e_{1}\rangle,\quad\hat{T}_{12}|e_{2}\rangle=\hat{T}_{34}|e_{2}\rangle=0,\\
\hat{T}_{23}|e_{1}\rangle=\hat{T}_{41}|e_{1}\rangle & =\frac{1}{d}\left(|e_{1}\rangle+\epsilon\sqrt{d^{2}-1}|e_{2}\rangle\right),\\
\hat{T}_{23}|e_{2}\rangle=\hat{T}_{41}|e_{2}\rangle &
=\frac{\sqrt{d^{2}-1}}{d}(\epsilon|e_{1}\rangle+\sqrt{d^{2}-1}|e_{2}\rangle).\end{align}
where the parameter $d$ represents the values of a loop, i.e.,
$d=\sqrt{2}$ for $b_{\mbox{I}}$ in Eq. (\ref{eq:b_tilt}) and $d=2$
for $b_{\mbox{II}}$ at $q=1$ in Eq. (\ref{eq:type_2_b}). From Eqs.
(\ref{eq:operator_A}) and (\ref{eq:operator_B}), the matrices $A$
and $B$ take the form\begin{align*} A & =\left(\begin{array}{cc}
\left(\alpha+\alpha^{-1}\right)d & 0\\
0 & \alpha\end{array}\right),\quad\hat{A}\left(\begin{array}{c}
|e_{1}\rangle\\
|e_{2}\rangle\end{array}\right)=A\left(\begin{array}{c}
|e_{1}\rangle\\
|e_{2}\rangle\end{array}\right),\\
B & =\frac{1}{\alpha d}\left(\begin{array}{cc}
\left(1+\alpha^{2}d\right) & \sqrt{d^{2}-1}\\
\sqrt{d^{2}-1} & \alpha^{2}d+(d^{2}-1)\end{array}\right),\quad\hat{B}\left(\begin{array}{c}
|e_{1}\rangle\\
|e_{2}\rangle\end{array}\right)=B\left(\begin{array}{c}
|e_{1}\rangle\\
|e_{2}\rangle\end{array}\right),\end{align*} when $d=\sqrt{2}$,
$\alpha=e^{i3\pi/8}$, we have \begin{align}
A_{\mbox{I}}=e^{-i\pi/8}\left(\begin{array}{cc}
1 & 0\\
0 & i\end{array}\right), & \quad B_{\mbox{I}}=\frac{1}{\sqrt{2}}e^{i\pi/8}\left(\begin{array}{cc}
1 & -i\\
-i & 1\end{array}\right),\label{eq:AB_type1}\end{align} and for
$d=2$, $\alpha=i$, we obtain\begin{align}
A_{\mbox{II}}=\left(\begin{array}{cc}
-1 & 0\\
0 & 1\end{array}\right), & \quad B_{\mbox{II}}=-\frac{1}{2}\left(\begin{array}{cc}
1 & -\sqrt{3}\\
-\sqrt{3} & -1\end{array}\right).\label{eq:AB_type2}\end{align}
where an overall factor $i$ has been dropped. $A$ and $B$ in
(\ref{eq:AB_type1})-(\ref{eq:AB_type2}) satisfy (\ref{eq:ABA_BAB}).
For type I, it has been proved in Refs {\cite{10}} that the
corresponding $\mathscr A(u)$and $\mathscr B(u)$ satisfy YBE
($u=\tan\theta/2$) (also see below
(\ref{eq:Lorentz})):\begin{align*}
\mathscr{A}_{\mbox{I}}(u)\mathscr{B}_{\mbox{I}}(\frac{u+v}{1+uv})\mathscr{A}_{\mbox{I}}(v)
&
=\mathscr{B}_{\mbox{I}}(v)\mathscr{A}_{\mbox{I}}(\frac{u+v}{1+uv})\mathscr{B}_{\mbox{I}}(u).\end{align*}
\begin{align}
 & \mathscr{A}_{\mbox{I}}(u)=\rho(u)\left(\begin{array}{cc}
\frac{1-u^{2}+2i\epsilon u}{1-u^{2}-2i\epsilon u} & 0\\
0 & 1\end{array}\right),\label{eq:cal_a_2}\\
 & \mathscr{B}_{\mbox{I}}(u)=\frac{\rho(u)}{1-u^{2}+2i\epsilon u}\left(\begin{array}{cc}
1-u^{2} & 2i\epsilon u\\
2i\epsilon u &
1-u^{2}\end{array}\right),\label{eq:cal_b_2}\end{align}  It is
interesting that the velocity additivity obeys the Lorentz form
($c=1$). Since type I corresponds to anyonic picture with
two-components, we expect that the velocity additivity rule of two
anyons may not obey the Galileo formula.

For type I, the operator $\hat{T}$ acts on $\left|e_{1}\right\rangle
$ and $\left|e_{2}\right\rangle $. In terms of the usual spin basis
at $i$-th and $j$-th spaces, we find
\begin{align*} \hat{T}_{ij} &
=\sqrt{2}\left(\left|\psi_{ij}\left\rangle \right\langle
\psi_{ij}\right|+\left|\phi_{ij}\left\rangle \right\langle
\phi_{ij}\right|\right),\end{align*} where \begin{align*}
\left|\psi_{ij}\right\rangle  & =\frac{1}{\sqrt{2}}\left(|\underset{i}{\uparrow}\underset{j}{\uparrow}\rangle+e^{-i\alpha}|\underset{i}{\downarrow}\underset{j}{\downarrow}\rangle\right),\\
\left|\phi_{ij}\right\rangle  & =\frac{1}{\sqrt{2}}\left(|\underset{i}{\uparrow}\underset{j}{\downarrow}\rangle-i|\underset{i}{\downarrow}\underset{j}{\uparrow}\rangle\right).\end{align*}
Correspondingly,\begin{align*}
\left|e_{1}\right\rangle  & =\frac{1}{\sqrt{2}}\left(\left|\psi_{12}\right\rangle \left|\psi_{34}\right\rangle +\left|\phi_{12}\right\rangle \left|\phi_{34}\right\rangle \right),\\
\left|e_{2}\right\rangle  & =\frac{1}{\sqrt{2}}\Big[\left(1-i\epsilon e^{i\alpha}\right)\left|\psi_{23}\right\rangle \left|\psi_{41}\right\rangle \\
 & \quad-\left(1-i\epsilon e^{i\alpha}\right)\left|\phi_{23}\right\rangle \left|\phi_{41}\right\rangle -\left|e_{1}\right\rangle \Big].\end{align*}
Whereas for type II, we have\begin{align*} \left|e_{1}\right\rangle
& =\left|\psi_{12}\right\rangle \left|\psi_{34}\right\rangle
,\end{align*}
\begin{align*}
\uij & =|\underset{i}{\uparrow}\underset{j}{\uparrow}\rangle+e^{-i\alpha}|\underset{i}{\downarrow}\underset{j}{\downarrow}\rangle=\sqrt{2}\left|\psi_{ij}\right\rangle ,\quad(j=i+1),\\
\nij & =\sqrt{2}\left\langle \psi_{ij}\right|,\\
\Unij & =\hat{T}_{i}=2\left|\psi_{ij}\left\rangle \right\langle \psi_{ij}\right|,\quad(j=i+1).\end{align*}

\section{Unified form for both types I and II}

In Sec. \ref{sec:Two-dimensional-braiding-matrices} we have
confirmed there are two types of YBE and their corresponding
$2\times2$ braid relation matrices (BRM). In this section we shall
demonstrate that the two types of $2\times2$ BRM are nothing but
Wigner's D-functions with $j=1/2$. The two types of braiding
matrices have $2\times2$ matrix forms and the corresponding
$4\times4$ matrix forms. They obey the T-L algebra and can be
Yang-Baxterized to yield solution of YBE. For i.e., $2$-body
scattering matrix $\breve{R}(x)$ is the elements of the matrix
representation of operator $\hat{T}$.

Is there an uniformed expression for both type I and II? The answer
is yes. We shall confirm that the matrix forms of
(\ref{eq:AB_type1}) and (\ref{eq:AB_type2}) are nothing but the
Wigner $D(\theta,\varphi)$-function {\cite{20}} with special values.

If we consider a simple three dimensional rotation transformation
for a two state system, entangled states may be connected with
natural basis by BRM. Therefore, we choose the original basis as
natural basis $\left|1\right\rangle $ and $\left|2\right\rangle $
since every two uni-orthogonal basis will actually achieve the same
result. After the transformation, the basis
$\left|E_{1}\right\rangle $ and $\left|E_{2}\right\rangle $ would
change to\begin{align} \left(\begin{array}{c}
\left|E_{1}\right\rangle \\
\left|E_{2}\right\rangle \end{array}\right) & =D^{1/2}(\theta,\varphi)\left(\begin{array}{c}
\left|1\right\rangle \\
\left|2\right\rangle \end{array}\right)\label{eq:E1-E2}\\
 & =\left(\begin{array}{cc}
\cos\frac{\theta}{2} & -\sin\frac{\theta}{2}e^{-i\varphi}\\
\sin\frac{\theta}{2}e^{i\varphi} & \cos\frac{\theta}{2}\end{array}\right)\left(\begin{array}{c}
\left|1\right\rangle \\
\left|2\right\rangle \end{array}\right).\nonumber \end{align}
$D^{1/2}(\theta,\varphi)$ is the matrix form of Wigner's D-function
{\cite{20}} with $J=1/2$. Here the states $\left|1\right\rangle $
and $\left|2\right\rangle $ not only represent spin-up and
spin-down, but also any objective of two dimensional representation,
say, $\left|1\right\rangle =\left|e_{1}\right\rangle $,
$\left|2\right\rangle =\left|e_{2}\right\rangle $ or
$\left|1\right\rangle
=\left|\uparrow\uparrow\cdots\uparrow\right\rangle $,
$\left|2\right\rangle
=\left|\downarrow\downarrow\cdots\downarrow\right\rangle $, etc. The
D-function $D(\theta,\varphi)$ means a rotation of angle $\theta$
about the axis $\mathbf{m}$, which is determined by $\varphi$
($\mathbf{m}=(-\sin\varphi,\cos\varphi,0)$). The notations of the
matrix forms of $D^{J}(\theta,\varphi)$ will be given in appendix
\ref{sec:Matrix_form_of_D-functions}.

In Ref. {\cite{21}} it had been proven that if D-function satisfy
the braid relation\begin{align}
D(\theta,\varphi_{1})D(\theta,\varphi_{2})D(\theta,\varphi_{1}) &
=D(\theta,\varphi_{2})D(\theta,\varphi_{1})D(\theta,\varphi_{2}),\label{eq:ddd}\end{align}
then $\theta$ and $\varphi$ should obey the relation
{\cite{21}}\begin{align} \cos\varphi &
=\frac{\cos\theta}{1-\cos\theta},\label{eq:theta_phi}\end{align}
where $\varphi=\varphi_{2}-\varphi_{1}$. Because Eq.
(\ref{eq:theta_phi}) only depends on the relative difference of
$\varphi_{1}$ and $\varphi_{2}$, we can set $\varphi_{1}=0$ and
$\varphi_{2}=\varphi$ for simplicity. Under these notations we can
get\begin{align} A(\theta &
)=D(\theta,\varphi_{1}=0)=\left(\begin{array}{cc}
\cos\frac{\theta}{2} & -\sin\frac{\theta}{2}\\
\sin\frac{\theta}{2} & \cos\frac{\theta}{2}\end{array}\right),\label{eq:a_theta}\\
B(\theta) & =D(\theta,\varphi_{2}=\varphi)=\left(\begin{array}{cc}
\cos\frac{\theta}{2} & -\sin\frac{\theta}{2}e^{-i\varphi}\\
\sin\frac{\theta}{2}e^{i\varphi} &
\cos\frac{\theta}{2}\end{array}\right).\label{eq:b_theta}\end{align}
Clearly $A(\theta)$ and $B(\theta)$ satisfy braid relation for
arbitrary $\theta$:\begin{align} A(\theta)B(\theta)A(\theta) &
=B(\theta)A(\theta)B(\theta).\label{eq:ABA_BAB_theta}\end{align} It
is emphasized that two different $\varphi$'s specify $A(\theta)$ and
$B(\theta)$ satisfying Eq. (\ref{eq:ABA_BAB_theta}). A different
proof is given in appendix \ref{sec:Expressions-of-the}

To obtain Eqs.(\ref{eq:AB_type1}) and (\ref{eq:AB_type2}) from
(\ref{eq:a_theta}) and (\ref{eq:b_theta}), let (\ref{eq:a_theta})
and (\ref{eq:b_theta}) subject to the unitary
transformation\begin{align} VA(\theta)V^{\dagger} &
=\left(\begin{array}{cc}
e^{i\theta/2} & 0\\
0 & e^{-i\theta/2}\end{array}\right),\label{eq:A_prime}\end{align}
and \begin{align}
VB(\theta)V & ^{\dagger}=\left(\begin{array}{cc}
\cos\frac{\theta}{2}+i\sin\frac{\theta}{2}\cos\varphi & i\sin\varphi\sin\frac{\theta}{2}\\
i\sin\varphi\sin\frac{\theta}{2} & \cos\frac{\theta}{2}-i\sin\frac{\theta}{2}\cos\varphi\end{array}\right),\label{eq:B_prime}\end{align}
where $V=\frac{1}{\sqrt{2}}\left(\begin{array}{cc}
1 & i\\
i & 1\end{array}\right).$ Obviously, Eq. (\ref{eq:A_prime}) is the
consequence by setting $\varphi=0$ in Eq. (\ref{eq:B_prime}). The
braid relation (\ref{eq:ddd}) constrains $\theta$ and $\varphi$ to
obey (\ref{eq:theta_phi}). We take two possibilities:
\begin{enumerate}
\item $\varphi=\pi/2$, $\theta=-\pi/2$, Eqs. (\ref{eq:A_prime}) and (\ref{eq:B_prime})
become into $e^{-i\pi/4}\left(\begin{array}{cc}
1 & 0\\
0 & i\end{array}\right)$ and $\frac{1}{\sqrt{2}}\left(\begin{array}{cc}
1 & -i\\
-i & 1\end{array}\right)$ respectively. By adjusting the phase
factor, we obtain (\ref{eq:AB_type1}).
\item $\varphi=2\pi/3$, $\theta=\pi,$ Eqs. (\ref{eq:A_prime}) and (\ref{eq:B_prime})
become into $(-i)\left(\begin{array}{cc}
-1 & 0\\
0 & 1\end{array}\right)$ and $(-\frac{i}{2})\left(\begin{array}{cc}
1 & -\sqrt{3}\\
-\sqrt{3} & -1\end{array}\right)$, which transfer to
(\ref{eq:AB_type2}) by omitting the overall factor $(-i)$.
\end{enumerate}
Correspondingly, for type I, the $4\times4$ $\breve{R}$-matrix is
found to be
\begin{align*}
\breve{R}(\theta,\varphi) & =\left(\begin{array}{cccc}
\cos\theta & 0 & 0 & e^{-i\varphi}\sin\theta\\
0 & \cos\theta & \sin\theta & 0\\
0 & -\sin\theta & \cos\theta & 0\\
-e^{i\varphi}\sin\theta & 0 & 0 &
\cos\theta\end{array}\right)\end{align*}.

\section{\label{sec:Extremum-of-D-function}Extremum of D-function and $\ell_{1}$-norm}

We should emphasize that only two sets of $\theta$ and $\varphi$,
i.e., $\left\{ \theta=-\pi/2,\varphi=\pi/2\right\} $ and $\left\{
\theta=\pi,\varphi=2\pi/3\right\} $ have the ``real'' physical
meanings. Since, the $4\times4$ form, there are just two types of
matrices in physics, i.e., $\hat{T}_{\mbox{II}}$ and
$\hat{T}_{\mbox{I}}$, for the familiar $6$-vertex model and quantum
information (Bell states) that connect with BRM and YBE.

It is interesting to ask whether this result is accidental or has
principle behind. We want to answer this question by introducing the
concept of $\ell_{1}$-norm.

If we take the $\ell_{1}$-norm of the coefficients of the
decomposition of $\left|E_{1}\right\rangle $ and
$\left|E_{2}\right\rangle $ in (\ref{eq:E1-E2}), we have
\begin{align*}
f(\theta)= & \left|\cos\frac{\theta}{2}\right|+\left|-\sin\frac{\theta}{2}e^{-i\varphi}\right|\\
= &
\left|\cos\frac{\theta}{2}\right|+\left|\sin\frac{\theta}{2}e^{i\varphi}\right|=\left|\cos\frac{\theta}{2}\right|+\left|\sin\frac{\theta}{2}\right|,\end{align*}
The two basis satisfy the same relation for $J_{z}=\pm1/2$. If
$\theta$ is restricted in the field $[-\pi,\pi]$, then
$\cos\frac{\theta}{2}$ is always positive. Also
$\sin\frac{\theta}{2}\geqslant0$ if $0\leqslant\theta\leqslant\pi$,
and $\sin\frac{\theta}{2}<0$ if $-\pi\leqslant\theta<0$. Using these
results, we can easily calculate the maximum and minimum values of
$f(\theta)$ and the corresponding $\theta$. When $\theta\in[0,\pi]$,
$f(\theta)=\cos\frac{\theta}{2}+\sin\frac{\theta}{2}$, $f(\theta)$
takes maximum value when $\theta=\pi/2$ while it takes minimum value
when $\theta=0$, $\pi$. When $\theta\in[-\pi,0]$,
$f(\theta)=\cos\frac{\theta}{2}-\sin\frac{\theta}{2}$, $f(\theta)$
takes maximum value when $\theta=-\pi/2$, and takes minimum value
when $\theta=-\pi$. Overall, when $\theta=-\pi$, $0$, or $\pi$,
$f(\theta)$ is minimum and when $\theta=-\pi/2$ or $\pi/2$,
$f(\theta)$ is maximum. These results can be seen in figure
\ref{Flo:j12}.

\begin{figure}
\centering{}\includegraphics[width=0.8\columnwidth]{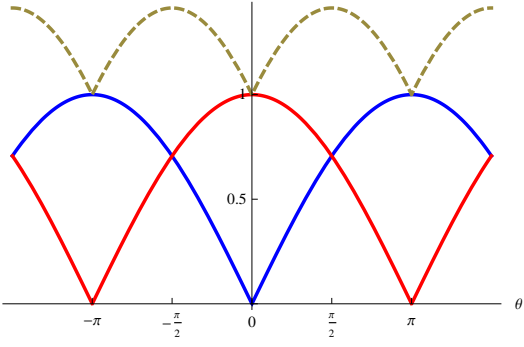}\caption{The blue and red lines  represent $\left|\cos\frac{\theta}{2}\right|$
and $\left|\sin\frac{\theta}{2}\right|$ separately, and the dashed
one indicates $\left|\cos\frac{\theta}{2}\right|+\left|\sin\frac{\theta}{2}\right|$
as for $J_{z}=\pm 1/2$. From the picture we can easily see the extremum
values of $f(\theta)$ are $\pm \pi$, $\pm \pi/2$ except the trivial value $\theta=0$. }
\label{Flo:j12}
\end{figure}

\subsection{\textmd{Type I BRM}}

By introducing the maximum of $\ell_{1}$-norm, if we choose
$\theta=-\pi/2$ and the corresponding $\varphi=\pi/2$ obtained by
Eq. (\ref{eq:theta_phi}), we get \begin{align*} A_{\mbox{I
}}=\frac{1}{\sqrt{2}}\left(\begin{array}{cc}
1 & 1\\
-1 & 1\end{array}\right),\quad & B_{\mbox{I }}=\frac{1}{\sqrt{2}}\left(\begin{array}{cc}
1 & -i\\
-i & 1\end{array}\right).\end{align*}

Braid relation will be still valid after a same constant unitary
transformation is acted on the matrices $A_{\mbox{I}}$ and
$B_{\mbox{I}}$. Using the unitary transformation
$V=\frac{1}{\sqrt{2}}\left(\begin{array}{cc}
1 & i\\
i & 1\end{array}\right)$ to make matrix $A_{\mbox{I}}$ diagonal:

\begin{align} A'_{\mbox{I}} &
=VA_{\mbox{I}}V^{\dagger}=e^{-i\pi/4}\left(\begin{array}{cc}
1 & 0\\
0 & i\end{array}\right),\label{eq:a_prime_1}\\
B'_{\mbox{I}} & =VB_{\mbox{I}}V^{\dagger}=\frac{1}{\sqrt{2}}\left(\begin{array}{cc}
1 & -i\\
-i & 1\end{array}\right),\label{eq:b_prime_1}\end{align} It is the
same as Eq. (\ref{eq:AB_type1}) except an overall phase factor.
Correspondingly, the YBE has the form\begin{align}
A(\theta_{1})B(\theta_{2},\varphi=\frac{\pi}{2})A(\theta_{3})=B(\theta_{3},\varphi=\frac{\pi}{2})A(\theta_{2})B(\theta_{1},\varphi=\frac{\pi}{2}),\label{eq:YBE1}\end{align}
and the spectral parameter $\theta$ should satisfy the
relation[10]\begin{align} \tan\frac{\theta_{2}}{2} &
=\frac{\tan\frac{\theta_{1}}{2}+\tan\frac{\theta_{3}}{2}}{1+\tan\frac{\theta_{1}}{2}\tan\frac{\theta_{3}}{2}}.\label{eq:Lorentz}\end{align}
By setting $u=\tan\frac{\theta}{2}$, this is just the additivity
rule of Lorentz velocity  ($c=1$).

\subsection{\textmd{Type II BRM}}

Now we substitute $\theta=\pi$ and corresponding $\varphi=2\pi/3$
which help the $\ell_{1}$-norm of $D^{1/2}(\theta,\varphi)$ to achieve
minimum, we obtain\begin{align*}
A_{\mbox{II }}=\frac{1}{\sqrt{2}}\left(\begin{array}{cc}
0 & -1\\
1 & 0\end{array}\right),\quad & B_{\mbox{II }}=\frac{1}{\sqrt{2}}\left(\begin{array}{cc}
0 & -e^{-2i\pi/3}\\
e^{2i\pi/3} & 0\end{array}\right).\end{align*} Taking the same
unitary transformation as for type I, we have\begin{align}
A'_{\mbox{II}} &
=VA_{\mbox{II}}V^{\dagger}=(-i)\left(\begin{array}{cc}
-1 & 0\\
0 & 1\end{array}\right),\label{eq:a_prime_2}\\
B'_{\mbox{II}} & =VB_{\mbox{II}}V^{\dagger}=(-\frac{i}{2})\left(\begin{array}{cc}
1 & -\sqrt{3}\\
-\sqrt{3} & -1\end{array}\right).\label{eq:b_prime_2}\end{align} The
same result is obtained as Eq. (\ref{eq:AB_type2}) except the
overall factor $(-i)$. The corresponding YBE relation
reads\begin{align}
A(\theta_{1})B(\theta_{2},\varphi=\frac{2\pi}{3})A(\theta_{3})=B(\theta_{3},\varphi=\frac{2\pi}{3})A(\theta_{2})B(\theta_{1},\varphi=\frac{2\pi}{3}),\label{eq:YBE2}\end{align}
and the spectral parameters should satisfy the relation for
$u=\tan\frac{\theta}{2}$\begin{align} u_{2} &
=u_{1}+u_{3}.\label{eq:Galileo-1}\end{align} This is just the
additivity rule of Galileo velocity .

When $\theta=0$, $A$ becomes a unit matrix, i.e., it is trivial. As
concerned to $\theta=-\pi$, $\pi/2$, we can substitute corresponding
$\varphi=-2\pi/3$, $-\pi/2$ into Eqs. (\ref{eq:a_theta}) and
 (\ref{eq:b_theta}). The results will just be the transposition of
the earlier matrices. If we change the order of original natural
basis and the entangled states basis, i.e., $(\left|2\right\rangle
,\left|1\right\rangle )$ and $(\left|E_{2}\right\rangle
,\left|E_{1}\right\rangle )$, the rotation transformation matrix
will also be the transposition of the original one.
\begin{align*} \left(\begin{array}{c}
\left|E_{2}\right\rangle \\
\left|E_{1}\right\rangle \end{array}\right) & =\left(\begin{array}{cc}
\cos\frac{\theta}{2} & -\sin\frac{\theta}{2}e^{-i\varphi}\\
\sin\frac{\theta}{2}e^{i\varphi} & \cos\frac{\theta}{2}\end{array}\right)^{T}\left(\begin{array}{c}
\left|2\right\rangle \\
\left|1\right\rangle \end{array}\right).\end{align*} Everything
changes to its transposition, they are still consistent.
Subsequently we just concentrate on the cases $\theta=\pi/2$ and
$\pi$.

In this way, we present that $\ell_{1}$-norm extremum can assist to
determine which $\theta$ and $\varphi$ have physical meanings.
Overall, the first type of BRM is related to the anyons and
entangled states, and the matrices are chosen by setting
$\theta=\pi/2$ and $\varphi=\pi/2$. The second type is connected to
fermions and bosons, and the BRM are chosen by setting $\theta=\pi$
and $\varphi=2\pi/3$. It is very interesting that the two types of
BRM, which really exist in physics, are just given by the extremum
of $\ell_{1}$-norm of the D-function:\begin{align*}
\mbox{maximizing} & \sum_{M=-1/2}^{1/2}\left|D_{MM'}^{1/2}(\theta,\varphi)\right|\mbox{ to get type I BRM},\\
\mbox{minimizing} & \sum_{M=-1/2}^{1/2}\left|D_{MM'}^{1/2}(\theta,\varphi)\right|\mbox{ to get type II BRM}.\end{align*}

In principle, the discussion for $j=1/2$ can be extended to any
dimensional spinor representations, see the appendix C.

\section{Motivation of using $\ell_{1}$-norm}

In our knowledge, up to now, there is no physical interpretation of
$\ell_{1}$-norm in QM, but in recent developments in the information
field, there has been strong motivation to take $\ell_{1}$-norm into
account.

There is a rapidly growing interest in the nonlinear sampling in
information theory, which is often referred to Compressive Sensing
(C-S) {\cite{22,23,24}}. It has had many applications to
information, digital sensors and computer tomography (CT) \cite{24}.
To explain C-S, let us consider a simple example. Suppose the
Fourier image of a signal $f(t)\ (t=n\frac{T}{N},n=1,2,\cdots,N)$ is
$\tilde{f}(\omega)=\sum_{i=1}^{k}\alpha_{i}\delta(\omega-\omega_{i})$.
If $k\ll N$, the signal is called ``sparse''. If a signal is sparse,
then much less measurements $y$ may be made to recover $f(t)=x$.
Suppose measuring matrix $\Phi$ is $M\times N$ matrix
($M\thickapprox k\log N\ll N$, where $k$ is ``sparsity'') , i.e.,
$y=\Phi x$. To recover $x$ ($N$ components), we can only measure $M$
data. Obviously, for given $y$ to find $x$ is an ill-posted problem
because $\Phi$ does not have the inverse. However, the C-S tells
that the recovery of $f(t)=x$ consists in {\cite{23}}
\begin{align} \mbox{minimize }\left\Vert x\right\Vert _{\ell_{1}} &
\quad\mbox{subject to }y=\Phi x.\label{eq:C-S}\end{align} The
$\ell_{1}$-norm plays the crucial role in Eq. (\ref{eq:C-S}).
Through this example, we can learn that the minimization of
$\ell_{1}$-norm can be used to determine some important physical
quantities. In Refs.{\cite{25,26}} the C-S theory has been used to
calculate density matrix. However, so far the concept of
$\ell_{1}$-norm is not emphasized in quantum information theory.

In Sec. \ref{sec:Extremum-of-D-function}, we have discussed one
possible usage of $\ell_{1}$-norm related to QM because of the
important application of $\ell_{1}$-norm in information theory. It
is reasonable to think there may be a deep connection between
$\ell_{1}$-norm and QM.

\section{Physical example related to YBE}

In appendix \ref{sec:Expressions-of-the} we have shown that the
matrix forms of the two types of 2D YBE are based on the basis
$\left|e_{1}\right\rangle $ and $\left|e_{2}\right\rangle $. With
this knowledge we can derive the basis of $A(\theta)$ and
$B(\theta)$, which satisfy\begin{align*} \left(\begin{array}{c}
\left|e_{1}'\right\rangle \\
\left|e_{2}'\right\rangle \end{array}\right) & =V\left(\begin{array}{c}
\left|e_{1}\right\rangle \\
\left|e_{2}\right\rangle \end{array}\right),\end{align*}
more specifically\begin{align*}
\left|e_{1}'\right\rangle  & =\frac{1}{\sqrt{2}}\left(\left|e_{1}\right\rangle +i\left|e_{2}\right\rangle \right),\\
\left|e_{2}'\right\rangle  &
=\frac{1}{\sqrt{2}}\left(i\left|e_{1}\right\rangle
+\left|e_{2}\right\rangle \right).\end{align*} After finding out
their connections with $\left|e_{1}\right\rangle $ and
$\left|e_{2}\right\rangle $, we can use the graph technique to show
the operators related to D-function. Here we confirm that
$\left|e_{1}'\right\rangle $ and $\left|e_{2}'\right\rangle $ are
nothing but the two basis of SU(2) algebra. Similar to atomic
physics, we define three operators: \begin{align*}
J_{+}= & \left|e_{1}'\right\rangle \left\langle e_{2}'\right|,\quad J_{-}=\left|e_{2}'\right\rangle \left\langle e_{1}'\right|,\\
J_{z}= & \frac{1}{2}\left(\left|e_{1}'\right\rangle \left\langle e_{1}'\right|-\left|e_{2}'\right\rangle \left\langle e_{2}'\right|\right).\end{align*}
Representing the operators as graph\begin{align*}
J_{+} & =\frac{1}{2}\left(\left|e_{1}\right\rangle +i\left|e_{2}\right\rangle \right)\left(i\left|e_{1}\right\rangle +\left|e_{2}\right\rangle \right)\\
 & =\frac{1}{2}\Big[\left(\left|e_{2}\right\rangle \left\langle e_{1}\right|+\left|e_{1}\right\rangle \left\langle e_{2}\right|\right)\\
 & \quad+i\left(-\left|e_{1}\right\rangle \left\langle e_{1}\right|+\left|e_{2}\right\rangle \left\langle e_{2}\right|\right)\Big]\\
 & =\frac{1}{2}\Bigg[\frac{1}{d\sqrt{d^{2}-1}}\left(\mn+\nm-\frac{2}{d}\mm\right)\\
 & \quad+\frac{i}{d^{2}-1}\left(\nn-\frac{1}{d}\mn-\frac{1}{d}\nm+\frac{-d^{2}+2}{d^{2}}\mm\right)\Bigg],\end{align*}
\begin{align*}
J_{-} & =\frac{1}{2}\Big[\left(\left|e_{2}\right\rangle \left\langle e_{1}\right|+\left|e_{1}\right\rangle \left\langle e_{2}\right|\right)\\
 & \quad+i\left(\left|e_{1}\right\rangle \left\langle e_{1}\right|-\left|e_{2}\right\rangle \left\langle e_{2}\right|\right)\Big]\\
 & =\frac{1}{2}\Bigg[\frac{1}{d\sqrt{d^{2}-1}}\left(\mn+\nm-\frac{2}{d}\mm\right)\\
 & \quad-\frac{i}{d^{2}-1}\left(\nn-\frac{1}{d}\mn-\frac{1}{d}\nm+\frac{-d^{2}+2}{d^{2}}\mm\right)\Bigg],\end{align*}
\begin{align*}
J_{z} & =\frac{i}{2}\left(\left|e_{2}\right\rangle \left\langle e_{1}\right|-\left|e_{1}\right\rangle \left\langle e_{2}\right|\right)\\
 & =\frac{i}{2d\sqrt{d^{2}-1}}\left(\nm-\mn\right).\end{align*}
Using graph technique, we can verify that \begin{align*}
\left[J_{z},\ J_{\pm}\right] & =\pm J_{\pm},\\
\left[J_{+},\ J_{-}\right] & =2J_{z}.\end{align*}
Noting\begin{align*}
J^{2} & =\frac{1}{2}\left(J_{+}J_{-}+J_{-}J_{+}\right)+J_{z}J_{z}\\
 & =\frac{3}{4}\left(\left|e_{1}'\right\rangle \left\langle e_{1}'\right|+\left|e_{2}'\right\rangle \left\langle e_{2}'\right|\right)\\
 & =\frac{3}{4}\left(\left|e_{1}\right\rangle \left\langle e_{1}\right|+\left|e_{2}\right\rangle \left\langle e_{2}\right|\right)\\
 & =\frac{3}{4}\frac{1}{d^{2}-1}\left(\nn-\frac{1}{d}\mn-\frac{1}{d}\nm+\mm\right),\end{align*}
it is easy to prove that \begin{align*} J^{2} &
\left|e_{i}'\right\rangle =\frac{3}{4}\left|e_{i}'\right\rangle
.\end{align*} Also we have\begin{align*}
\left(J_{+}\right)^{2} & =\left(J_{-}\right)^{2}=0,\\
J_{z}\left|e_{1}'\right\rangle  & =\frac{1}{2}\left|e_{1}'\right\rangle ,\\
J_{z}\left|e_{2}'\right\rangle  &
=-\frac{1}{2}\left|e_{2}'\right\rangle .\end{align*} These
calculations let us see that $\left|e_{1}'\right\rangle $ and
$\left|e_{2}'\right\rangle $ are nothing but the two basis of SU(2)
algebra.

The 2-states can be understood in terms of the Cooper pair of
superconductivity. Through the mean field approximation, the
four-fermion interaction\begin{align*} H &
=\sum_{\vec{k}}\varepsilon_{\vec{k}}a_{\vec{k}}^{+}a_{\vec{k}}+\sum_{\sigma=\pm}A_{\vec{k}\vec{k}'}a_{-\vec{k}\sigma}^{+}a_{\vec{k}\sigma}^{+}a_{\vec{k}\sigma}a_{-\vec{k}\sigma}\end{align*}
reduces to \begin{align*} H_{0} &
=\sum_{\vec{k}}H_{\vec{k}}^{0},\end{align*} where\begin{align*}
H_{\vec{k}}^{0} &
=\varepsilon_{\vec{k}}J_{z}^{\vec{k}}+\frac{1}{2}\Delta_{\vec{k}}^{*}J_{-}^{\vec{k}}+\frac{1}{2}\Delta_{\vec{k}}^{*}J_{+}^{\vec{k}},\end{align*}
$\Delta_{\vec{k}}=\sum_{\vec{k}'}A_{\vec{k}\vec{k}'}\left\langle
a_{\vec{k}'\downarrow},a_{-\vec{k}\uparrow}\right\rangle $ and
$J_{\pm}^{\vec{k}}$, $J_{z}^{\vec{k}}$ satisfy the SU(2)-algebra.
The operator $D(\xi)$ can be used to diagonalize the $H_{\vec{k}}$
for a fixed $\vec{k}$:\begin{align*}
D(\xi)=e^{\xi J_{+}-\xi^{*}J_{-}} & =e^{\tau J_{+}}e^{\ln(1+|\tau|^{2})J_{z}}e^{-\tau^{*}J_{-}},\\
D(\xi)H_{\vec{k}}D^{\dagger}(\xi)=E_{\vec{k}}J_{z}, & \quad E_{\vec{k}}=\left(\varepsilon_{\vec{k}}^{2}+\left|\Delta_{\vec{k}}\right|^{2}\right)^{1/2},\\
\Delta_{\vec{k}}=\left|\Delta_{\vec{k}}\right|e^{i\varphi/2}, & \quad\tan\theta_{\vec{k}}=\frac{\left|\Delta_{\vec{k}}\right|}{\varepsilon_{\vec{k}}},\\
\xi=\frac{\theta}{2}e^{-i\varphi}, &
\quad\tau=e^{-i\varphi}\tan\frac{\theta}{2}.\end{align*} where
$J_{+}$, $J_{-}$ and $J_{z}$ are angular momentum operators. In
terms of the fermion operators\begin{align*}
J_{\vec{k}}^{+}=a_{-\vec{k}\downarrow}^{+} & a_{\vec{k}\uparrow}^{+},\quad J_{\vec{k}}^{-}=a_{\vec{k}\uparrow}a_{-\vec{k}\downarrow},\\
J_{\vec{k}}^{z}=\frac{1}{2} &
\left(n_{\vec{k}\uparrow}+n_{-\vec{k}\downarrow}-1\right),\end{align*}
and defining\begin{align*} J_{-}^{\vec{k}} & \left|0\right\rangle
_{\vec{k}}=0\quad\left|\xi\right\rangle _{\vec{k}}=e^{\xi
J_{+}^{k}-\xi*J_{-}^{k}}\left|0,0\right\rangle
_{\vec{k}}\end{align*} where $\left|0,0\right\rangle _{\vec{k}}$ is
the eigen state of
$\left|n_{\vec{k}\uparrow}=0,n_{-\vec{k}\downarrow}=0\right\rangle
$, then \begin{align*}
\left|\xi\right\rangle _{\vec{k}} & =e^{\tau J_{+}}e^{\ln(1+|\tau|^{2})J_{z}}e^{-\tau^{*}J_{-}}\left|0,0\right\rangle _{\vec{k}}\\
 & =e^{\tau J_{+}}e^{\ln(1+|\tau|^{2})J_{z}}\left|0,0\right\rangle _{\vec{k}}\\
 & =\frac{1}{\sqrt{1+|\tau|^{2}}}\left(\left|0,0\right\rangle _{\vec{k}}+\tau\left|1,1\right\rangle _{\vec{k}}\right),\end{align*}
where $\left|0,0\right\rangle
_{k}=|n_{\vec{k}\uparrow}=0,n_{-\vec{k}\downarrow}=0\rangle$ and
$\left|1,1\right\rangle
_{k}=|n_{\vec{k}\uparrow}=1,n_{-\vec{k}\downarrow}=1\rangle$, i.e.,
$J_{z}|0,0\rangle=-\frac{1}{2}|0,0\rangle$ and
$J_{z}|1,1\rangle=+\frac{1}{2}|1,1\rangle$. It is easy to find
\begin{align*} D(\xi)\left(\begin{array}{c}
\left|0,0\right\rangle \\
\left|1,1\right\rangle \end{array}\right) & =\left(\begin{array}{cc}
\cos\frac{\theta}{2} & \sin\frac{\theta}{2}e^{-i\varphi}\\
-\sin\frac{\theta}{2}e^{i\varphi} &
\cos\frac{\theta}{2}\end{array}\right)\left(\begin{array}{c}
\left|0,0\right\rangle \\
\left|1,1\right\rangle \end{array}\right),\end{align*} namely
$|0,0\rangle$ and $|1,1\rangle$ can be served as $|2\rangle$ and
$|1\rangle$ in section \ref{sec:Extremum-of-D-function}. Suppose
$\theta$ and $\varphi$ are taken to be $(-\pi/2,\pi/2)$ and
$(\pi,2\pi/3)$, respectively. It yields the same matrix as given by
(\ref{eq:AB_type1}). The ground energy $E_{\vec{k}}$ degenerates to
$\varphi$, which can be detected through Josephson current. With
this sense, CS may be served as simulation of YBE for any
$\Delta_{\vec{k}}$, i.e., $\theta_{\vec{k}}$ with the corresponding
$\varphi_{\vec{k}}$.

\section{Examples of $J=1$ and $J=3/2$}

In appendix \ref{sec:Extremum-points-of} it has been provided
evidence that for arbitrary $j=1/2,1,3/2,\cdots$ etc.,
$\sum_{M'=-J}^{M'=J}\left|D_{MM'}^{J}(\theta,\varphi)\right|$ can
reach its extreme value when $\theta=-\pi,-\pi/2,0,\pi/2,\pi$
($\theta\in[-\pi,\pi]$). This result can generalize the result for
$J=1/2$ and the corresponding $2\times2$ BRM. In this section we
shall calculate the $\ell_{1}$-norm of $D_{MM'}^{J}(\theta,\varphi)$
for $J=1$ and $J=3/2$, and demonstrate that the extremum of
$\ell_{1}$-norm lead to the two types of BRM.

\subsection{$J=1$}

We take the $\ell_{1}$-norm of every row of the D-function
$D_{MM'}^{1}(\theta,\varphi)$. From Eqs. (\ref{eq:a1}) and
(\ref{eq:b1}), it can be derived that
$\left|D_{MM'}^{1}(\theta,\varphi)\right|=\left|d_{MM'}^{1}(\theta)\right|$,
and the first and third row share the same results, therefor we just
concentrate on the first two rows. We can prove that for
$\theta\in[-\pi,\pi]$, $\ell_{1}$-norm can achieve its extremum
value when $\theta=-\pi,-\pi/2,0,\pi/2,\pi$. For detailed
calculations, please refer to appendix \ref{sec:Extremum-points-of}.
The maximum and minimum can be seen easily from pictures
\ref{Flo:j1a}
and \ref{Flo:j1b}.%
\begin{figure}[h]
\begin{centering}
\includegraphics[width=0.85\columnwidth]{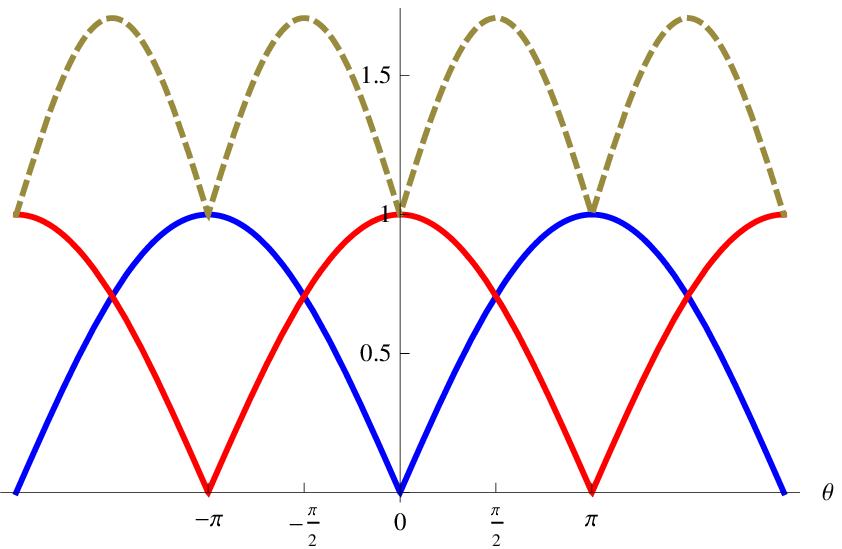}
\par\end{centering}

\begin{centering}
\caption{The blue and red lines represent $\left|\cos\frac{\theta}{2}\right|$
and $\left|\sin\frac{\theta}{2}\right|$ separately, and the dashed
one indicates $f(\theta)_{1}^{1}=\left|(1+\cos\theta)/2\right|+\left|-\sin\theta/\sqrt{2}\right|+\left|(1-\cos\theta)/2\right|$,
as for $J_{z}=\pm1$ states. From the picture we can easily see the
extremum values of $f(\theta)_{1}^{1}$ and the corresponding $\theta$. }
\label{Flo:j1a}
\par\end{centering}

\end{figure}

\begin{figure}[h]
\begin{centering}
\includegraphics[width=0.85\columnwidth]{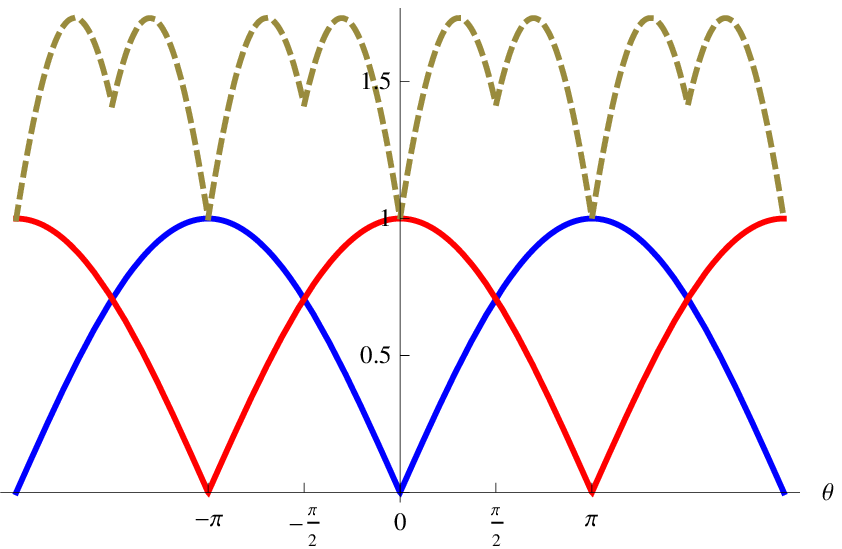}
\par\end{centering}

\centering{}\caption{The blue and red lines represent $\left|\cos\frac{\theta}{2}\right|$
and $\left|\sin\frac{\theta}{2}\right|$ separately, and the dashed
one indicates $f(\theta)_{2}^{1}=\left|\sin\theta/\sqrt{2}\right|+\left|\cos\theta\right|+\left|-\sin\theta/\sqrt{2}\right|$
as for $J_{z}=0$ state. From the picture we can easily see the extremum
values of $f(\theta)_{2}^{1}$ and the corresponding $\theta$. }
\label{Flo:j1b}
\end{figure}

Although Fig. (\ref{Flo:j1b}) has more extremum points than Fig.
(\ref{Flo:j1a}), there are just five points that figures
\ref{Flo:j1a} and  \ref{Flo:j1b} share together. From the view of
point in Fig. (\ref{Flo:j1a}), we can still choose $\theta=\pi/2$
and $\pi$ to find out two types of BRM.
\begin{enumerate}
\item Substituting $\theta=\pi/2$ and $\varphi=\pi/2$ into Eqs. (\ref{eq:a1})
and (\ref{eq:b1}), we get \begin{align*}
A_{\mbox{I}}^{1}=\left(\begin{array}{ccc}
\frac{1}{2} & -\frac{1}{\sqrt{2}} & \frac{1}{2}\\
\frac{1}{\sqrt{2}} & 0 & -\frac{1}{\sqrt{2}}\\
\frac{1}{2} & \frac{1}{\sqrt{2}} & \frac{1}{2}\end{array}\right),\quad & B_{\mbox{I}}^{1}=\left(\begin{array}{ccc}
\frac{1}{2} & \frac{i}{\sqrt{2}} & -\frac{1}{2}\\
\frac{i}{\sqrt{2}} & 0 & \frac{i}{\sqrt{2}}\\
-\frac{1}{2} & \frac{i}{\sqrt{2}} & \frac{1}{2}\end{array}\right).\end{align*}
After taking the unitary transformation\begin{align*}
\widetilde{A_{\mbox{I}}^{1}} & =T^{\dagger}A_{\mbox{I}}^{1}T=\left(\begin{array}{ccc}
-i & 0 & 0\\
0 & 1 & 0\\
0 & 0 & i\end{array}\right),\\
\widetilde{B_{\mbox{I}}^{1}} & =T^{\dagger}B_{\mbox{I}}^{1}T=\left(\begin{array}{ccc}
\frac{1}{2} & \frac{1}{\sqrt{2}} & \frac{1}{2}\\
-\frac{1}{\sqrt{2}} & 0 & \frac{1}{\sqrt{2}}\\
\frac{1}{2} & -\frac{1}{\sqrt{2}} & \frac{1}{2}\end{array}\right),\end{align*}
where $T=\left(\begin{array}{ccc}
\frac{1}{2} & \frac{1}{\sqrt{2}} & \frac{1}{2}\\
\frac{i}{\sqrt{2}} & 0 & -\frac{i}{\sqrt{2}}\\
-\frac{1}{2} & \frac{1}{\sqrt{2}} & -\frac{1}{2}\end{array}\right).$ This is the $3\times3$ type I BRM.
\item Substituting $\theta=\pi$ and $\varphi=2\pi/3$ into Eqs. (\ref{eq:a1})
and (\ref{eq:b1}), we get\begin{align*}
A_{\mbox{II}}^{1}=\left(\begin{array}{ccc}
0 & 0 & 1\\
0 & -1 & 0\\
1 & 0 & 0\end{array}\right),\quad & B_{\mbox{II}}^{1}=\left(\begin{array}{ccc}
0 & 0 & e^{2i\pi/3}\\
0 & -1 & 0\\
e^{-2i\pi/3} & 0 & 0\end{array}\right).\end{align*}
After taking the same unitary transformation\begin{align*}
\widetilde{A_{\mbox{II}}^{1}} & =\left(\begin{array}{ccc}
-1 & 0 & 0\\
0 & 1 & 0\\
0 & 0 & -1\end{array}\right),\quad\widetilde{B_{\mbox{II}}^{1}}=\left(\begin{array}{ccc}
-\frac{1}{4} & i\frac{\sqrt{6}}{4} & \frac{3}{4}\\
-i\frac{\sqrt{6}}{4} & -\frac{1}{2} & -i\frac{\sqrt{6}}{4}\\
\frac{3}{4} & i\frac{\sqrt{6}}{4} & -\frac{1}{4}\end{array}\right).\end{align*}
This is the $3\times3$ type II BRM.
\end{enumerate}
Both types of BRM satisfy the braid relation $\widetilde{A^{1}}\widetilde{B^{1}}\widetilde{A^{1}}=\widetilde{B^{1}}\widetilde{A^{1}}\widetilde{B^{1}}$.
It should be noted that for vector solutions ($J=1,2,\cdots$) when
$\theta=\pm\pi/2$, the $\ell_{1}$-norm of D-function matrices may
achieve minimum value, not maximum as the case for spinor solutions
($J=1/2$, $3/2$, $\cdots$).

A valuable attention is that for $J=1$ the maximum at $\pm\pi/2$ and
minimums at $\pm\pi$ for states $J_{z}=\pm1$ are the same, but for
$J_{z}=0$ state the $\pm\pi/2$ are the minimum. This state should be
singleted, the physical interpretation is chiral photon. This
picture does not occur in spinors, i.e., for $J$ is half integers.

\subsection{$J=3/2$}

We take the $\ell_{1}$-norm of every row of the D-function
$D_{MM'}^{3/2}(\theta,\varphi)$. From Eqs. (\ref{eq:a32}) and
(\ref{eq:b32}) it can be derived that
$\left|D_{MM'}^{3/2}(\theta,\varphi)\right|=\left|d_{MM'}^{3/2}(\theta)\right|$,
also the first and third row have the same results. We just
concentrate on the first two rows. We can also prove that for
$\theta\in[-\pi,\pi]$, $\ell_{1}$-norm can achieve its extremum
value when $\theta=-\pi,-\pi/2,0,\pi/2,\pi$. The maximum and minimum
can be seen easily from pictures \ref{Flo:j32a} and \ref{Flo:j32b}.

\begin{figure}[h]
\centering{}\includegraphics[width=0.85\columnwidth]{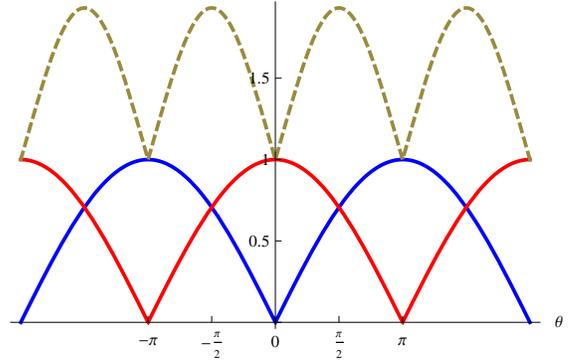}\caption{The blue and red lines  represent $\left|\cos\frac{\theta}{2}\right|$
and $\left|\sin\frac{\theta}{2}\right|$ separately, and the dashed
one indicates $f(\theta)_{1}^{3/2}=\left|\cos^{3}\frac{\theta}{2}\right|+\left|-\sqrt{3}\sin\frac{\theta}{2}\cos^{2}\frac{\theta}{2}\right|+\left|\sqrt{3}\sin^{2}\frac{\theta}{2}\cos\frac{\theta}{2}\right|+\left|-\sin^{3}\frac{\theta}{2}\right|$
as for $J_{z}=\pm3/2$. From the picture we can easily see the extremum
values of $f(\theta)_{1}^{1}$ and the corresponding $\theta$. }
\label{Flo:j32a}
\end{figure}

\begin{figure}[h]
\centering{}\includegraphics[width=0.85\columnwidth]{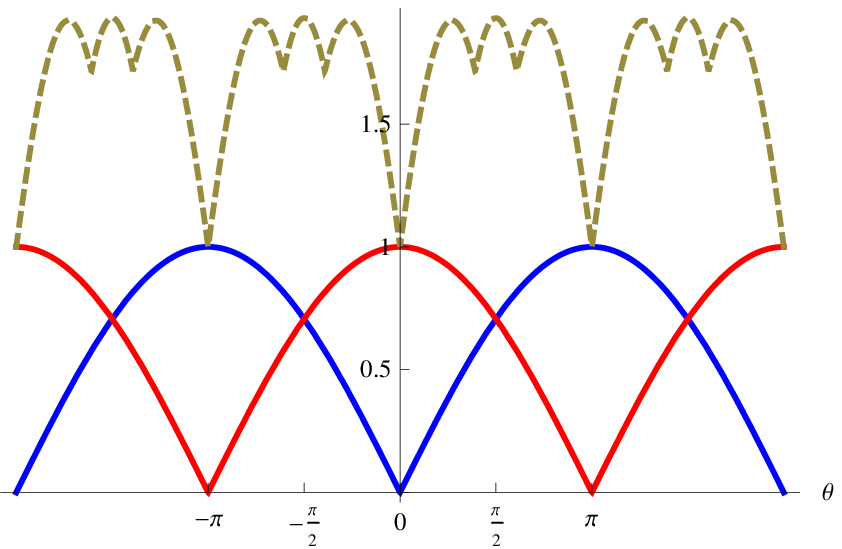}\caption{The blue and red lines represent $\left|\cos\frac{\theta}{2}\right|$
and $\left|\sin\frac{\theta}{2}\right|$ separately, and the dashed
one indicates $f(\theta)_{2}^{3/2}=\left|\sqrt{3}\sin\frac{\theta}{2}\cos^{2}\frac{\theta}{2}\right|+\left|\cos\frac{\theta}{2}\left(3\cos^{2}\frac{\theta}{2}-2\right)\right|+\left|\sin\frac{\theta}{2}\left(3\sin^{2}\frac{\theta}{2}-2\right)\right|+\left|\sqrt{3}\sin^{2}\frac{\theta}{2}\cos\frac{\theta}{2}\right|$
as for $J_{z}=\pm1/2$. From the picture we can easily see the extremum
values of $f(\theta)_{2}^{1}$ and the corresponding $\theta$. }
\label{Flo:j32b}
\end{figure}
Although Fig. (\ref{Flo:j32b}) has more extremum points than Fig.
(\ref{Flo:j32a}), there are just five points that they share
together. From the view of point in Fig. (\ref{Flo:j32a}), we can
choose $\theta=\pi/2$ and $\pi$ to find the two types of BRM.
\begin{enumerate}
\item Substituting $\theta=\pi/2$ and $\varphi=\pi/2$ into Eqs. (\ref{eq:a32})
and (\ref{eq:b32}), we get \begin{align*} A_{\mbox{I}}^{3/2} &
=\left(\begin{array}{cccc}
\frac{1}{2\sqrt{2}} & -\frac{\sqrt{6}}{4} & \frac{\sqrt{6}}{4} & -\frac{1}{2\sqrt{2}}\\
\frac{\sqrt{6}}{4} & -\frac{1}{2\sqrt{2}} & -\frac{1}{2\sqrt{2}} & \frac{\sqrt{6}}{4}\\
\frac{\sqrt{6}}{4} & \frac{1}{2\sqrt{2}} & -\frac{1}{2\sqrt{2}} & -\frac{\sqrt{6}}{4}\\
\frac{1}{2\sqrt{2}} & \frac{\sqrt{6}}{4} & \frac{\sqrt{6}}{4} & \frac{1}{2\sqrt{2}}\end{array}\right),\\
B_{\mbox{I}}^{3/2} & =\left(\begin{array}{cccc}
\frac{1}{2\sqrt{2}} & i\frac{\sqrt{6}}{4} & -\frac{\sqrt{6}}{4} & -\frac{i}{2\sqrt{2}}\\
i\frac{\sqrt{6}}{4} & -\frac{1}{2\sqrt{2}} & \frac{i}{2\sqrt{2}} & -\frac{\sqrt{6}}{4}\\
-\frac{\sqrt{6}}{4} & \frac{i}{2\sqrt{2}} & -\frac{1}{2\sqrt{2}} & i\frac{\sqrt{6}}{4}\\
-\frac{i}{2\sqrt{2}} & -\frac{\sqrt{6}}{4} & i\frac{\sqrt{6}}{4} & \frac{1}{2\sqrt{2}}\end{array}\right).\end{align*}
After taking the unitary transformation\begin{align*}
\widetilde{A_{\mbox{I}}^{3/2}} & =U^{\dagger}A_{\mbox{I}}^{1}U=\left(\begin{array}{cccc}
-e^{i\pi/4} & 0 & 0 & 0\\
0 & e^{i\pi/4} & 0 & 0\\
0 & 0 & e^{-i\pi/4} & 0\\
0 & 0 & 0 & -e^{-i\pi/4}\end{array}\right),\\
\widetilde{B_{\mbox{I}}^{3/2}} & =U^{\dagger}B_{\mbox{I}}^{1}U=\left(\begin{array}{cccc}
\frac{1}{2\sqrt{2}} & \frac{\sqrt{6}}{4} & -\frac{\sqrt{6}}{4} & -\frac{1}{2\sqrt{2}}\\
\frac{\sqrt{6}}{4} & -\frac{1}{2\sqrt{2}} & \frac{1}{2\sqrt{2}} & -\frac{\sqrt{6}}{4}\\
\frac{\sqrt{6}}{4} & -\frac{1}{2\sqrt{2}} & -\frac{1}{2\sqrt{2}} & \frac{\sqrt{6}}{4}\\
\frac{1}{2\sqrt{2}} & \frac{\sqrt{6}}{4} & \frac{\sqrt{6}}{4} & \frac{1}{2\sqrt{2}}\end{array}\right),\end{align*}
where $U=\left(\begin{array}{cccc}
\frac{i}{2\sqrt{2}} & i\frac{\sqrt{6}}{4} & -i\frac{\sqrt{6}}{4} & -\frac{i}{2\sqrt{2}}\\
-\frac{\sqrt{6}}{4} & \frac{1}{2\sqrt{2}} & \frac{1}{2\sqrt{2}} & -\frac{\sqrt{6}}{4}\\
-i\frac{\sqrt{6}}{4} & \frac{i}{2\sqrt{2}} & -\frac{i}{2\sqrt{2}} & i\frac{\sqrt{6}}{4}\\
\frac{1}{2\sqrt{2}} & \frac{\sqrt{6}}{4} & \frac{\sqrt{6}}{4} & \frac{1}{2\sqrt{2}}\end{array}\right).$ This is the $4\times4$ type I BRM.
\item Substituting $\theta=\pi$ and $\varphi=2\pi/3$ into Eqs. (\ref{eq:a32})
and (\ref{eq:b32}), we get\begin{align*} A_{\mbox{II}}^{3/2} &
=\left(\begin{array}{cccc}
0 & 0 & 0 & -1\\
0 & 0 & 1 & 0\\
0 & -1 & 0 & 0\\
1 & 0 & 0 & 0\end{array}\right),\\
B_{\mbox{II}}^{3/2} & =\left(\begin{array}{cccc}
0 & 0 & 0 & -1\\
0 & 0 & e^{-2i\pi/3} & 0\\
0 & -e^{2i\pi/3} & 0 & 0\\
1 & 0 & 0 & 0\end{array}\right).\end{align*}
After taking the same unitary transformation\begin{align*}
\widetilde{A_{\mbox{II}}^{3/2}} & =\left(\begin{array}{cccc}
i & 0 & 0 & 0\\
0 & i & 0 & 0\\
0 & 0 & -i & 0\\
0 & 0 & 0 & -i\end{array}\right),\\
\widetilde{B_{\mbox{II}}^{3/2}} & =\left(\begin{array}{cccc}
-\frac{i}{8} & i\frac{3\sqrt{3}}{8} & \frac{3}{8} & -\frac{3\sqrt{3}}{8}\\
i\frac{3\sqrt{3}}{8} & i\frac{5}{8} & -\frac{\sqrt{3}}{8} & \frac{3}{8}\\
-\frac{3}{8} & \frac{\sqrt{3}}{8} & -i\frac{5}{8} & -i\frac{3\sqrt{3}}{8}\\
\frac{3\sqrt{3}}{8} & -\frac{3}{8} & -i\frac{3\sqrt{3}}{8} & \frac{i}{8}\end{array}\right).\end{align*}
This is the $4\times4$ type II BRM.
\end{enumerate}
Both types of BRM satisfy the braid relation $\widetilde{A^{3/2}}\widetilde{B^{3/2}}\widetilde{A^{3/2}}=\widetilde{B^{3/2}}\widetilde{A^{3/2}}\widetilde{B^{3/2}}$.

The physical interpretation of the $3\times3$ and $4\times4$ BRM
remains to be discovered. We want to emphasize that based on the
proof in appendix \ref{sec:Extremum-points-of} we can further
calculate $n\times n$ BRM ($n=2J+1$ is an arbitrary integer). In
this section, we just give the two simplest examples $n=3$ and $4$
to show the difference between spinors and vectors.

\section{Conclusion}

In quantum mechanics, we should normalize a wave function, so we are
familiar with $\ell_{2}$-norm but not $\ell_{1}$-norm. Considering
the important application of the $\ell_{1}$-norm theory in the
information theory, we try to introduce the $\ell_{1}$-norm to QM
through the three-dimensional rotation transformation for spin
system. It turns out that by taking the extremum of $\ell_{1}$-norm
of D-functions with $J=1/2$, we can derive the two types of YBE,
which have important physical interpretation. One of them is
connected with anyons and entangled states while the other is
related to the usual low dimensional integrable models.

By the end, we generalize the result to the D-functions with$j=1$,
$3/2$ and find out that they have the same property. This result
shows there may be a deep connection between $\ell_{1}$-norm and QM,
D-functions as well as YBE. The same properties are held for any J
being half integers, see appendix C. However, extending the
discussions for 2 x 2 and 4 x 4 braiding matrices to any J is a
challenge problem.

\section*{Acknowledgment}

We thank Prof. Guang-Hong Chen, Prof. Z. H. Wang and Dr. Xu-Biao
Peng for their helpful discussions. A special appreciation to Prof
Z.H.Wang for his beautiful lectures at Chern Institute. This work
was supported in part by SRFDP (200800550015) and Liu- Hui Center of
Nankai University and Tianjin University. Additional support was
provided by the Ministry of Science and Technology of China
(2009IM033000).

\appendix

\section{Expressions of the two types of YBE\label{sec:Expressions-of-the}}

By using the same calculation method, we can obtain the two explicit
types of 2D YBE. We set
$\check{R}_{ii+1}(u)=a_{i}(u)I+b_{i}(u)T_{ii+1}$ and act $\breve{R}$
on $\left|e_{1}\right\rangle $ and $\left|e_{2}\right\rangle $; then
we have\begin{align}
\check{R}_{12}(u)|e_{1}\rangle & =[a_{1}(u)+db_{1}(u)]|e_{1}\rangle,\\
\check{R}_{12}(u)|e_{2}\rangle & =a_{1}(u)|e_{1}\rangle,\\
\check{R}_{23}(u)|e_{1}\rangle & =\left[a_{2}(u)+\frac{b_{2}(u)}{d}\right]|e_{1}\rangle+\epsilon\frac{\sqrt{d^{2}-1}}{d}b_{2}(u)|e_{2}\rangle,\\
\check{R}_{23}(u)|e_{2}\rangle &
=\epsilon\frac{\sqrt{d^{2}-1}}{d}b_{2}(u)|e_{1}\rangle+\left[a_{2}(u)+\frac{d^{2}-1}{d}b_{2}(u)\right]|e_{2}\rangle.\end{align}
These should give the 2-D representations of $\check{R}_{12}(u)$ and
$\check{R}_{23}(u)$. Defining \begin{eqnarray}
\begin{array}{l}
\mathscr{A}_{ij}(u)=\langle e_{i}|\check{R}_{12}(u)|e_{j}\rangle,\\
\mathscr{B}_{ij}(u)=\langle e_{i}|\check{R}_{23}(u)|e_{j}\rangle,\ \ (i,j=1,2).\end{array}\end{eqnarray}
 we have \begin{eqnarray}
\mathscr{A}(u)=\left(\begin{array}{cc}
a_{1}(u)+db_{1}(u) & 0\\
0 & a_{1}(u)\end{array}\right),\end{eqnarray}
 \begin{align}
\mathscr{B}(u)=\left(\begin{array}{cc}
a_{2}(u)+\frac{b_{2}(u)}{d} & \epsilon\frac{\sqrt{d^{2}-1}}{d}b_{2}(u)\\
\epsilon\frac{\sqrt{d^{2}-1}}{d}b_{2}(u) & a_{2}(u)+\frac{d^{2}-1}{d}b_{2}(u)\end{array}\right).\end{align}
The YBE should be satisfied for the $1$-D momentum conservation:
\begin{eqnarray}
\mathscr{A}(u)\mathscr{B}(u+v)\mathscr{A}(v)=\mathscr{B}(v)\mathscr{A}(u+v)\mathscr{B}(u).\label{eq:Galileo}\end{eqnarray}
 To simplify the independent relations, we take the special case where
\begin{eqnarray}
a_{1}(u)=a_{2}(u)=a(u),\ \ b_{1}(u)=b_{2}(u)=b(u).\end{eqnarray}
The only constraint equation is simplified to

\begin{align*}
 & [a(u)b(v)+b(u)a(v)+db(v)b(u)]a(u+v)\\
 & =[a(v)a(u)-b(u)b(v)]b(u+v).\end{align*}
 Setting \begin{align*}
a(u) & =\rho(u),\quad b(u)=\rho(u)G(u),\end{align*}
 the YBE leads to \begin{eqnarray}
G(u)=\frac{u}{\gamma-u},\end{eqnarray}
 for $d=2$ ($\gamma$ is arbitrary), \begin{align}
 & \mathscr{A}(u)=\rho(u)\left(\begin{array}{cc}
\frac{\gamma+u}{\gamma-u} & 0\\
0 & 1\end{array}\right),\label{eq:cal_a}\\
 & \mathscr{B}(u)=\frac{\rho(u)}{2(\gamma-u)}\left(\begin{array}{cc}
2\gamma-u & \epsilon\sqrt{3}u\\
\epsilon\sqrt{3}u &
2\gamma+u\end{array}\right).\label{eq:cal_b}\end{align} This is the
second type of 2D YBE, which satisfies the Galileo velocity addition
rule as shown in Eq. (\ref{eq:Galileo-1}). If we introduce the
transformation\begin{align*} \frac{\gamma+u}{\gamma-u}\equiv
e^{-i\theta}, & \quad\rho(u)\equiv e^{i\theta/2},\end{align*}
then\begin{align*}
\rho(u)\frac{u}{\gamma-u}=-i\sin\frac{\theta}{2},\quad &
\rho(u)\frac{\gamma}{\gamma-u}=\cos\frac{\theta}{2}.\end{align*} We
can use these notations to obtain the following
matrices\begin{align*} \mathscr{A}(u) & =\left(\begin{array}{cc}
e^{-i\theta/2} & 0\\
0 & e^{i\theta/2}\end{array}\right)\equiv A'(\theta),\\
\mathscr{B}(u) & =\left(\begin{array}{cc}
\cos\frac{\theta}{2}+\frac{i}{2}\sin\frac{\theta}{2} & i\frac{\sqrt{3}}{2}\sin\frac{\theta}{2}\\
i\frac{\sqrt{3}}{2}\sin\frac{\theta}{2} &
\cos\frac{\theta}{2}-\frac{i}{2}\sin\frac{\theta}{2}\end{array}\right)\equiv
B'(\theta,\varphi=\frac{2\pi}{3}).\end{align*} By using this
notations we have identified the expressions of (\ref{eq:A_prime})
(\ref{eq:B_prime}) and (\ref{eq:cal_a}) (\ref{eq:cal_b}) except
$\theta\to-\theta$.

From Ref. {\cite{10}}, the first type of 2D YBE, we introduce the
transformation\begin{align*} \frac{1+\beta^{2}u^{2}+2i\epsilon\beta
u}{1+\beta^{2}u^{2}-2i\epsilon\beta u}\equiv e^{-i\theta}, &
\quad\rho(u)\equiv e^{-i\theta/2},\end{align*} We then obtain the
following matrices\begin{align*} \mathscr{A}(u) &
=\left(\begin{array}{cc}
e^{-i\theta/2} & 0\\
0 & e^{i\theta/2}\end{array}\right)\equiv A'(\theta),\\
\mathscr{B}(u) & =\left(\begin{array}{cc}
\cos\frac{\theta}{2} & -i\sin\frac{\theta}{2}\\
-i\sin\frac{\theta}{2} &
\cos\frac{\theta}{2}\end{array}\right)\equiv
B'(\theta,\varphi=\frac{\pi}{2}).\end{align*} In this way we
identify the expressions of (\ref{eq:A_prime}) (\ref{eq:B_prime})
and (\ref{eq:cal_a_2}) (\ref{eq:cal_b_2}) except $\theta\to-\theta$.
From the above expressions we see clearly that the matrix form of
D-function gives a uniform way to describe the two types of 2D YBE.

\section{\label{sec:Matrix_form_of_D-functions}Matrix form of D-functions}

\subsection{Notations}

The general D-function expression is {\cite{20}} \begin{align*}
D(\alpha,\beta,\gamma) & =e^{-i\alpha J_{z}}e^{-i\beta
J_{y}}e^{-i\gamma J_{z}}.\end{align*} However, the D-function we
used in this paper has a special property, following Perelomov
{\cite{27}}\begin{align*} D(\theta,\varphi) &
=e^{-i\theta\mathbf{m\cdot J}}=e^{\zeta
J_{+}-\zeta*J_{-}},\end{align*} where \begin{align}
\zeta=-\frac{\theta}{2}e^{-i\varphi}, &
\quad\mathbf{m}=(-\sin\varphi,\cos\varphi,0),\label{eq:zeta}\end{align}
$\mathbf{J}$ is the angular momentum operator. There is another form
of D-function\begin{align}
D(n) & =e^{nJ_{+}}e^{\ln(1+|n|^{2})J_{z}}e^{-n^{*}J_{-}},\label{eq:d_n}\\
 & n=-\tan\frac{\theta}{2}e^{-i\varphi}\label{eq:n}\end{align}

The D-function $D(\theta,\varphi)$ means a rotation of angle
$\theta$ about the axis $\mathbf{m}$ which is determined by
$\varphi$ as shown in Eq.(\ref{eq:zeta}). This specific operator
$D(\theta,\varphi)$ was used to generate spin coherent states
{\cite{27,28}}.

It is easy to calculate the relation between $D(\alpha,\beta,\gamma)$
and $D(\theta,\varphi)$, i.e.,\begin{align*}
\theta & =\beta,\\
\varphi & =\alpha=-\gamma,\end{align*}
so\begin{align*}
D(\theta,\varphi) & =e^{-i\varphi J_{z}}e^{-\theta J_{y}}e^{i\varphi Jz}.\end{align*}

Then the matrix form of the rotation operator $D(\theta,\varphi)$
would be \begin{align}
D_{MM'}^{J}(\theta,\varphi) & =\left\langle J,M\right|e^{-i\varphi J_{z}}e^{-\theta J_{y}}e^{i\varphi Jz}\left|J,M'\right\rangle \nonumber \\
 & =e^{-i\varphi M}e^{i\varphi M'}\left\langle J,M\right|e^{-\theta J_{y}}\left|J,M'\right\rangle \nonumber \\
 & =e^{i\varphi(-M+M')}d_{MM'}^{J}(\theta).\label{eq:Dd}\end{align}

We need to let matrix $A$ be the same as $d_{MM'}^{J}(\theta)$ and
matrix $B$ as $D_{MM'}^{J}(\theta,\varphi)$. It indicates that $A$
comes from a rotation of angle $\theta$ along the $y$ axis and $B$
comes from the same rotation angle, but along the axis of
$(-\sin\varphi,\cos\varphi,0)$, which can be obtained by rotating
$y$ of the angle $\varphi$ along $z$ axis.

By choosing $\varphi=0$ in Eq. (\ref{eq:Dd}), we get \begin{align} A
&
=D_{MM'}^{J}(\theta,\varphi=0)=d_{MM'}^{J}(\theta),\label{eq:a}\end{align}
so $B$ satisfies \begin{align} B= &
D_{MM'}^{J}(\theta,\varphi)=e^{i\varphi(-M+M')}d_{MM'}^{J}(\theta).\label{eq:b}\end{align}

In Ref. \cite{20}, the explicit forms of $d_{MM'}^{J}(\theta)$ had
been given.

\subsection{Example $J=1/2$}

The values of $d_{MM'}^{1/2}\left(\theta\right)$ are shown in table
\ref{Flo:d12}:

\begin{center}
\begin{table}[H]
\begin{centering}
\begin{tabular}{ccc}
\hline
\backslashbox{$M$}{$M'$} & $\frac{1}{2}$ & $-\frac{1}{2}$\tabularnewline
\hline
\noalign{\vskip\doublerulesep}
$\frac{1}{2}$ & $\cos\frac{\theta}{2}$ & $-\sin\frac{\theta}{2}$\tabularnewline[\doublerulesep]
\noalign{\vskip\doublerulesep}
\noalign{\vskip\doublerulesep}
$-\frac{1}{2}$ & $\sin\frac{\theta}{2}$ & $\cos\frac{\theta}{2}$\tabularnewline[\doublerulesep]
\hline
\noalign{\vskip\doublerulesep}
\end{tabular}
\par\end{centering}

\caption{Explicit form of $d_{MM'}^{1/2}\left(\theta\right)$}
\label{Flo:d12}
\end{table}

\par\end{center}

Then\begin{align}
A & =d_{MM'}^{1/2}\left(\theta\right)=\left(\begin{array}{cc}
\cos\frac{\theta}{2} & -\sin\frac{\theta}{2}\\
\sin\frac{\theta}{2} & \cos\frac{\theta}{2}\end{array}\right).\label{eq:a12}\end{align}
\begin{align}
B & =D_{MM'}^{1/2}\left(\theta,\varphi\right)=A\odot\left(\begin{array}{cc}
1 & e^{-i\varphi}\\
e^{i\varphi} & 1\end{array}\right)\nonumber \\
 & =\left(\begin{array}{cc}
\cos\frac{\theta}{2} & -\sin\frac{\theta}{2}e^{-i\varphi}\\
\sin\frac{\theta}{2}e^{i\varphi} & \cos\frac{\theta}{2}\end{array}\right),\label{eq:b12}\end{align}
where $\odot$ means entrywise product.

\subsection{Example $J=1$}

For $J=1$, following the same procedure, we get

\begin{align}
A & =d_{MM'}^{1}\left(\theta\right)=\left(\begin{array}{ccc}
\frac{1+\cos\theta}{2} & -\frac{\sin\theta}{\sqrt{2}} & \frac{1-\cos\theta}{2}\\
\frac{\sin\theta}{\sqrt{2}} & \cos\theta & -\frac{\sin\theta}{\sqrt{2}}\\
\frac{1-\cos\theta}{2} & \frac{\sin\theta}{\sqrt{2}} & \frac{1+\cos\theta}{2}\end{array}\right),\label{eq:a1}\end{align}
\begin{align}
B & =D_{MM'}^{1}\left(\theta,\varphi\right)=A\odot\left(\begin{array}{ccc}
1 & e^{-i\varphi} & e^{-2i\varphi}\\
e^{i\varphi} & 1 & e^{-i\varphi}\\
e^{2i\varphi} & e^{i\varphi} & 1\end{array}\right)\nonumber \\
 & =\left(\begin{array}{ccc}
\frac{1+\cos\theta}{2} & -\frac{\sin\theta}{\sqrt{2}}e^{-i\varphi} & \frac{1-\cos\theta}{2}e^{-2i\varphi}\\
\frac{\sin\theta}{\sqrt{2}}e^{i\varphi} & \cos\theta & -\frac{\sin\theta}{\sqrt{2}}e^{-i\varphi}\\
\frac{1-\cos\theta}{2}e^{2i\varphi} & \frac{\sin\theta}{\sqrt{2}}e^{i\varphi} & \frac{1+\cos\theta}{2}\end{array}\right).\label{eq:b1}\end{align}

\subsection{Example $J=3/2$}

For $J=3/2$, by following the same procedure, we obtain

\begin{widetext}\begin{align}
A & =d_{MM'}^{3/2}\left(\theta\right)=\left(\begin{array}{cccc}
\cos^{3}\frac{\theta}{2} & -\sqrt{3}\sin\frac{\theta}{2}\cos^{2}\frac{\theta}{2} & \sqrt{3}\sin^{2}\frac{\theta}{2}\cos\frac{\theta}{2} & -\sin^{3}\frac{\theta}{2}\\
\sqrt{3}\sin\frac{\theta}{2}\cos^{2}\frac{\theta}{2} & \cos\frac{\theta}{2}\left(3\cos^{2}\frac{\theta}{2}-2\right) & \sin\frac{\theta}{2}\left(3\sin^{2}\frac{\theta}{2}-2\right) & \sqrt{3}\sin^{2}\frac{\theta}{2}\cos\frac{\theta}{2}\\
\sqrt{3}\sin^{2}\frac{\theta}{2}\cos\frac{\theta}{2} & -\sin\frac{\theta}{2}\left(3\sin^{2}\frac{\theta}{2}-2\right) & \cos\frac{\theta}{2}\left(3\cos^{2}\frac{\theta}{2}-2\right) & -\sqrt{3}\sin\frac{\theta}{2}\cos^{2}\frac{\theta}{2}\\
\sin^{3}\frac{\theta}{2} & \sqrt{3}\sin^{2}\frac{\theta}{2}\cos\frac{\theta}{2} & \sqrt{3}\sin\frac{\theta}{2}\cos^{2}\frac{\theta}{2} & \cos^{3}\frac{\theta}{2}\end{array}\right),\label{eq:a32}\end{align}

\begin{align}
B & =D_{MM'}^{3/2}\left(\theta,\varphi\right)=\left(\begin{array}{cccc}
\cos^{3}\frac{\theta}{2} & -\sqrt{3}\sin\frac{\theta}{2}\cos^{2}\frac{\theta}{2}e^{-i\varphi} & \sqrt{3}\sin^{2}\frac{\theta}{2}\cos\frac{\theta}{2}e^{-2i\varphi} & -\sin^{3}\frac{\theta}{2}e^{-3i\varphi}\\
\sqrt{3}\sin\frac{\theta}{2}\cos^{2}\frac{\theta}{2}e^{i\varphi} & \cos\frac{\theta}{2}\left(3\cos^{2}\frac{\theta}{2}-2\right) & \sin\frac{\beta}{2}\left(3\sin^{2}\frac{\theta}{2}-2\right)e^{-i\varphi} & \sqrt{3}\sin^{2}\frac{\theta}{2}\cos\frac{\theta}{2}e^{-2i\varphi}\\
\sqrt{3}\sin^{2}\frac{\theta}{2}\cos\frac{\theta}{2}e^{2i\varphi} & -\sin\frac{\theta}{2}\left(3\sin^{2}\frac{\theta}{2}-2\right)e^{i\varphi} & \cos\frac{\theta}{2}\left(3\cos^{2}\frac{\theta}{2}-2\right) & -\sqrt{3}\sin\frac{\theta}{2}\cos^{2}\frac{\theta}{2}e^{-i\varphi}\\
\sin^{3}\frac{\theta}{2}e^{3i\varphi} & \sqrt{3}\sin^{2}\frac{\theta}{2}\cos\frac{\theta}{2}e^{2i\varphi} & \sqrt{3}\sin\frac{\theta}{2}\cos^{2}\frac{\theta}{2}e^{i\varphi} & \cos^{3}\frac{\theta}{2}\end{array}\right).\label{eq:b32}\end{align}

\end{widetext}

\section{Extremum points of the $\ell_{1}$-norm of D-function\label{sec:Extremum-points-of}}

We shall prove for arbitrary $j=1/2,1,3/2,\cdots$ etc.,
$\sum_{M'=-J}^{M'=J}\left|D_{MM'}^{J}(\theta,\varphi)\right|$ can
reach its extreme value when $\theta=-\pi,-\pi/2,0,\pi/2,\pi$
($\theta\in[-\pi,\pi]$). From Eq. (\ref{eq:Dd}) we know
$\left|D_{MM'}^{J}(\theta,\varphi)\right|=\left|d_{MM'}^{J}(\theta)\right|$,
therefore from now on, we focus on calculating
$\sum_{M'=-J}^{M'=J}\left|d_{MM'}^{J}(\theta)\right|$.

\subsection{General results}

The explicit expression of D-function $d_{MM'}^{J}(\theta)$ is
\cite{20}:

\begin{align}
d_{MM'}^{J}(\theta) & =\left[(J+M)!(J-M)!(J+M')!(J-M')!\right]^{1/2}\nonumber \\
 & \quad\times\sum_{\chi}\frac{(-1)^{\chi}}{(J-M-\chi)!(J+M'-\chi)!(\chi+M-M')!\chi!}\label{eq:d}\\
 & \quad\times\left(\cos\frac{\theta}{2}\right)^{2J+M'-M-2\chi}\left(-\sin\left(\frac{\theta}{2}\right)\right)^{M-M'+2\chi},\nonumber \end{align}
 where $\chi$ is arbitrary integer. In our case, we need to fix
$J$, $M$ and take the sum of $M'$. So $J-M=a$ is a constant.
Substituting $M=J-a$ into Eq. (\ref{eq:d})  \begin{align}
d_{J-a\ M'}^{J}(\theta) & =\left[(2J-a)!a!(J+M')!(J-M')!\right]^{1/2}\nonumber \\
 & \quad\times\sum_{\chi}\frac{(-1)^{\chi}}{(a-\chi)!(J+M'-\chi)!(\chi+J-a-M')!\chi!}\nonumber \\
 & \quad\times\left(\cos\frac{\theta}{2}\right)^{J+a+M'-2\chi}\left(-\sin\left(\frac{\theta}{2}\right)\right)^{J-a-M'+2\chi}.\label{eq:d_general}\end{align}

From the equation $\frac{1}{n!}=0$ (if $n<0$), we can derive the
relational expression $\chi$ should satisfy \begin{align*}
0\leqslant\chi\leqslant J-M=a.\end{align*} Letting\begin{align}
A & =\left[(2J-a)!a!(J+M')!(J-M')!\right]^{1/2},\label{eq:A_a}\\
B(\chi) & =\frac{(-1)^{\chi}}{(a-\chi)!(J+M'-\chi)!(\chi+J-a-M')!\chi!},\label{eq:bx}\end{align}
we have

\begin{align}
\frac{\partial}{\partial\theta}d_{J-a\ M'}^{J} & =A\cdot\sum_{\chi}B(\chi)\cdot\frac{1}{2}\left(\cos\frac{\theta}{2}\right)^{J+a+M'-2\chi-1}\nonumber \\
 & \quad\cdot\left(-\sin\left(\frac{\theta}{2}\right)\right)^{J-a-M'+2\chi-1}\\
 & \quad\times\Big[(J+a+M'-2\chi)\left(-\sin\left(\frac{\theta}{2}\right)\right)^{2}\label{eq:partial_d}\\
 & \quad-(J-a-M'+2\chi)\left(\cos\frac{\theta}{2}\right)^{2}\Big].\end{align}

Now we shall prove the existence of the five extremum points one by
one. $\theta=\pi/2$ is our first consideration.

\subsection{$\theta=\pi/2$}

There are four steps to prove $\theta=\pi/2$ leads to extremum value
of $\sum_{M'=-J}^{M'=J}\left|d_{MM'}^{J}(\theta)\right|$.

\subsubsection{The relation of two D-function matrix elements}

At first, we need to proof\begin{align} d_{J-a\
M'}^{J}(\frac{\pi}{2})=\pm(-1)^{2M'}d_{J-a\
-M'}^{J}(\frac{\pi}{2}).\label{eq:d_d_2}\end{align}

Substitute $\theta=\frac{\pi}{2}$ into Eq. (\ref{eq:d_general})\begin{align*}
d_{J-a\ -M'}^{J}(\frac{\pi}{2}) & =\left[(2J-a)!a!(J-M')!(J+M')!\right]^{1/2}\\
 & \quad\times\sum_{\chi=0}^{\chi=a}\frac{(-1)^{\chi}}{(a-\chi)!(J-M'-\chi)!(\chi+J-a+M')!\chi!}\\
 & \quad\times\left(\cos\left(\frac{\pi}{4}\right)\right)^{J+a-M'-2\chi}\left(-\sin\left(\frac{\pi}{4}\right)\right)^{J-a+M'+2\chi}\\
 & =A\left(\frac{1}{\sqrt{2}}\right)^{2J}(-1)^{J-a+M'+2\chi}\\
 & \quad\cdot\sum_{\chi=0}^{\chi=a}\frac{(-1)^{\chi}}{(a-\chi)!(J-M'-\chi)!(\chi+J-a+M')!\chi!}\\
\text{letting } & \chi'=a-\chi\\
 & =A\left(\frac{1}{\sqrt{2}}\right)^{2J}(-1)^{J+a+M'-2\chi'}\\
 & \quad\cdot\sum_{\chi'=a}^{\chi'=0}\frac{(-1)^{a-\chi'}}{\chi'!(\chi'+J-a-M')!(J+M'-\chi')!(a-\chi')!}\\
 & =\pm A\left(\frac{1}{\sqrt{2}}\right)^{2j}(-1)^{J-a-M'+2\chi}\\
 & \quad\cdot\sum_{\chi=a}^{\chi=0}\frac{(-1)^{\chi}}{\chi!(\chi+J-a-M')!(j+M'-\chi)!(a-\chi)!}\\
 & =\pm(-1)^{2M'}d_{J-a\ M'}^{J}.\end{align*}

Its sign ($+$ or $-$) is determined by the parities of $\chi$ and
$a-\chi$. If they have the same parities, it is $+$ sign, and $a$ is
even in this situation. On the other hand, it is $-$ sign and $a$ is
odd. From now on wherever there are two signs, we just mark the
upper one work for even $a$ and the lower one work for odd $a$.

\subsubsection{The derivation of the two D-function matrix elements}

Second we proof\begin{align}
\left.\frac{\partial}{\partial\theta}d_{J-a\ M'}^{J}(\theta)\right|_{\theta=\frac{\pi}{2}}=\mp(-1)^{2M'}\left.\frac{\partial}{\partial\theta}d_{J-a\ -M'}^{J}(\theta)\right|_{\theta=\frac{\pi}{2}}.\label{eq:pd_pd_2}\end{align}
When $\theta=\frac{\pi}{2}$\begin{align}
\left.\frac{\partial}{\partial\theta}d_{J-a\ M'}^{J}(\theta)\right|_{\theta=\frac{\pi}{2}} & =A\frac{1}{2}\left(\frac{1}{\sqrt{2}}\right)^{2J-2}(-1)^{J-a-M'+1}\nonumber \\
 & \quad\cdot\sum_{\chi=0}^{a}B(\chi)\left[a+M'-2\chi\right].\label{eq:bPi2}\end{align}

Let $C(M',\chi)=B(\chi)\left[a+M'-2\chi\right]$, it is easy to proof
that if we set $M'\to-M'$ and $\chi\to a-\chi$, then\begin{align*}
C(-M',a-\chi) & =\frac{(-1)^{a-\chi}}{(a-\chi)!\chi!}\cdot\frac{\left[2\chi-a-m'\right]}{(\chi+j-m'-a)!(j+m'-\chi)}\\
 & =\pm\frac{(-1)^{\chi}}{(a-\chi)!\chi!}\cdot\frac{\left[2\chi-a-m'\right]}{(\chi+j-m'-a)!(j+m'-\chi)}\\
 & =\mp C(M',\chi)\end{align*}
So Eq. (\ref{eq:bPi2}) can be written as\begin{align*}
\left.\frac{\partial}{\partial\theta}d_{J-a\ -M'}^{J}(\theta)\right|_{\theta=\frac{\pi}{2}} & =A\frac{1}{2}\left(\frac{1}{\sqrt{2}}\right)^{2J-2}(-1)^{J-a+M'+1}\\
 & \quad\cdot\sum_{\chi=0}^{\chi=a}B(\chi)\left[a-M'-2\chi\right]\\
 & =A\frac{1}{2}\left(\frac{1}{\sqrt{2}}\right)^{2J-2}(-1)^{J-a+M'+1}\cdot\sum_{\chi=0}^{\chi=a}C(-M',\chi)\\
\text{\ensuremath{\text{letting }}} & \chi'=a-\chi\\
 & =A\frac{1}{2}\left(\frac{1}{\sqrt{2}}\right)^{2J-2}(-1)^{J-a+M'+1}\cdot\sum_{\chi'=a}^{\chi'=0}C(-M',a-\chi')\\
\mbox{because } & C(-M',a-\chi)=\mp C(m',\chi)\\
 & =\mp A\frac{1}{2}\left(\frac{1}{\sqrt{2}}\right)^{2J-2}(-1)^{J-a-M'+1}(-1)^{2M'}\\
 & \quad\cdot\sum_{\chi'=a}^{\chi'=0}C(M',\chi')\\
 & =\mp(-1)^{2M'}A\frac{1}{2}\left(\frac{1}{\sqrt{2}}\right)^{2j-2}(-1)^{j-a-M'+1}\\
 & \quad\cdot\sum_{\chi=a}^{\chi=0}B(\chi)\left[a+m'-2\chi\right]\\
\mbox{because } & A\mbox{ is is symmetric to }M'\text{and}-M'\\
 & =\mp(-1)^{2M'}\left.\frac{\partial}{\partial\beta}d_{j-a\ m'}^{j}(\beta)\right|_{\beta=\frac{\pi}{2}}.\end{align*}

\subsubsection{The $\ell_{1}$-norm of the two D-function matrix elements}

Next we proof when $\theta=\frac{\pi}{2}$, $\left|d_{J-a\
M'}^{J}(\theta)\right|+\left|d_{J-a\ -M'}^{J}(\theta)\right|$
reaches its extreme value, the mathematical expression
is\begin{align}
\left.\frac{\partial}{\partial\theta}\left(\left|d_{J-a\
M'}^{J}(\theta)\right|+\left|d_{J-a\
-M'}^{J}(\theta)\right|\right)\right|_{\theta=\frac{\pi}{2}}=0.\label{eq:absabs}\end{align}
We shall prove it in the following different situations.
\begin{enumerate}
\item $d_{J-a\ M'}^{J}(\frac{\pi}{2})=\pm(-1)^{2M'}d_{J-a\ -M'}^{J}(\frac{\pi}{2})=0$.
Modulus has the relation $\left|d_{J-a\ M'}^{J}(\theta)\right|+\left|d_{J-a\ -M'}^{J}(\theta)\right|\geqslant0$,
it is easy to see that $\left|d_{J-a\ M'}^{J}(\frac{\pi}{2})\right|+\left|d_{J-a\ -M'}^{J}(\frac{\pi}{2})\right|=0$
is the extreme values.
\item $d_{J-a\ M'}^{J}(\frac{\pi}{2})=\pm(-1)^{2M'}d_{J-a\ -M'}^{J}(\frac{\pi}{2})\neq0$.
For fixed $J,a,M'$, the function $d_{J-a\ M'}^{J}(\theta)$ is
infinite-order differentiable. From Eq. (\ref{eq:d_d_2}), if
$d_{J-a\ M'}^{J}(\frac{\pi}{2})$ is not equal to zero, then $d_{J-a\
-M'}^{J}(\frac{\pi}{2})$ neither and vice versa. So there is an
epsilon neighborhood of $\theta=\frac{\pi}{2}$ where $d_{J-a\
M'}^{J}(\theta)$ and $d_{J-a\ -M'}^{J}(\theta)$ are not equal to
zero, then \begin{align*}
 & \left.\frac{\partial}{\partial\theta}\left(\left|d_{J-a\ M'}^{J}(\theta)\right|+\left|d_{J-a\ -M'}^{J}(\theta)\right|\right)\right|_{\theta=\frac{\pi}{2}}\\
= & \left.\frac{\partial}{\partial\theta}\left|\left(d_{J-a\
M'}^{J}(\theta)\pm(-1)^{2M'}d_{J-a\
-M'}^{J}(\theta)\right)\right|\right|_{\theta=\frac{\pi}{2}}.\end{align*}
From Eq. (\ref{eq:pd_pd_2}), we have \[
\frac{\partial}{\partial\theta}\left.\left(d_{J-a\
M'}^{J}(\theta)\pm(-1)^{2M'}d_{J-a\
-M'}^{J}(\theta)\right)\right|_{\theta=\frac{\pi}{2}}=0,\] so in the
end \begin{align*}
\left.\frac{\partial}{\partial\theta}\left(\left|d_{J-a\
M'}^{J}(\theta)\right|+\left|d_{J-a\
-M'}^{J}(\theta)\right|\right)\right|_{\theta=\frac{\pi}{2}}=0.\end{align*}

\end{enumerate}

\subsubsection{Final proof}

In the final step,we proof $\sum_{M'=-J}^{M'=J}\left|d_{MM'}^{J}(\theta)\right|$
reaches its extreme value when $\theta=\frac{\pi}{2}$:\begin{align*}
\frac{\partial}{\partial\theta}\left.\left(\sum_{M'=-J}^{M'=J}\left|d_{MM'}^{J}(\theta)\right|\right)\right|_{\theta=\frac{\pi}{2}}=0.\end{align*}
It is easy to see\begin{align*}
 & \frac{\partial}{\partial\theta}\left.\left(\sum_{M'=-J}^{M'=J}\left|d_{MM'}^{J}(\theta)\right|\right)\right|_{\theta=\frac{\pi}{2}}\\
= & \sum_{M'=-J}^{M'=-1/2}\left.\frac{\partial}{\partial\theta}\left(\left|d_{J-a\ M'}^{J}(\theta)\right|+\left|d_{J-a\ -M'}^{J}(\theta)\right|\right)\right|_{\theta=\frac{\pi}{2}}\\
 & +\left.\frac{\partial}{\partial\theta}\left|d_{J-a\ 0}^{J}(\theta)\right|\right|_{\theta=\frac{\pi}{2}}\\
\text{combining with } & \mbox{Eq.(\ref{eq:absabs})}\\
= & 0.\end{align*}
$\left.\frac{\partial}{\partial\theta}\left|d_{J-a\
0}^{J}(\theta)\right|\right|_{\theta=\frac{\pi}{2}}$ only exists
when $2J$ is even, and
$\left.\frac{\partial}{\partial\theta}\left|d_{J-a\
0}^{J}(\theta)\right|\right|_{\theta=\frac{\pi}{2}}=0$ can be seen
easily from Eqs. (\ref{eq:d_d_2}) and (\ref{eq:pd_pd_2}). Finally we
verify that
$\sum_{M'=-J}^{M'=J}\left|d_{MM'}^{J}(\theta)\right|$reaches its
extreme value when $\theta=\frac{\pi}{2}$.

As for other extremum points, the proven methods are the same, only
the important calculation was demonstrated here.

\subsection{$\theta=-\pi/2$}

The method is the same as the one used in the previous section, so
only important calculations are shown here.

Substitute $\theta=-\frac{\pi}{2}$ into Eq. (\ref{eq:d_general})\begin{align*}
d_{J-a\ -M'}^{J}(-\frac{\pi}{2}) & =A\left(\frac{1}{\sqrt{2}}\right)^{2J}\\
 & \quad\cdot\sum_{\chi=0}^{\chi=a}\frac{(-1)^{\chi}}{(a-\chi)!(J-M'-\chi)!(\chi+J-a+M')!\chi!}\\
\text{letting } & \chi'=a-\chi\\
 & =\pm d_{J-a\ M'}^{J}(-\frac{\pi}{2}).\end{align*}

When $\theta=-\frac{\pi}{2}$ \begin{align}
\left.\frac{\partial}{\partial\theta}d_{J-a\ M'}^{J}(\theta)\right|_{\theta=-\frac{\pi}{2}} & =A\frac{1}{2}\left(\frac{1}{\sqrt{2}}\right)^{2J-2}\cdot\sum_{\chi=0}^{a}B(\chi)\left[a+M'-2\chi\right].\label{eq:b_Pi2}\end{align}

Eq. (\ref{eq:b_Pi2}) can be written as \begin{align*}
\left.\frac{\partial}{\partial\theta}d_{J-a\ -M'}^{J}(\theta)\right|_{\theta=-\frac{\pi}{2}} & =A\frac{1}{2}\left(\frac{1}{\sqrt{2}}\right)^{2j-2}\cdot\sum_{\chi=0}^{\chi=a}C(-M',\chi)\\
 & =\mp\left.\frac{\partial}{\partial\theta}d_{J-a\ M'}^{J}(\theta)\right|_{\theta=-\frac{\pi}{2}}.\end{align*}

\begin{enumerate}
\item $d_{J-a\ M'}^{J}(-\frac{\pi}{2})=\pm d_{J-a\ -M'}^{J}(-\frac{\pi}{2})=0$.
Modulus has the relation $\left|d_{J-a\ M'}^{J}(\theta)\right|+\left|d_{J-a\ -M'}^{J}(\theta)\right|\geqslant0$,
it is easy to see that $\left|d_{J-a\ M'}^{J}(-\frac{\pi}{2})\right|+\left|d_{J-a\ -M'}^{J}(-\frac{\pi}{2})\right|=0$
is the extreme values.
\item $d_{J-a\ M'}^{J}(-\frac{\pi}{2})=\pm d_{J-a\ -M'}^{J}(-\frac{\pi}{2})\neq0$.
For fixed $J,a,M'$, the function $d_{J-a\ M'}^{J}(\theta)$ is
infinite-order differentiable. Also we know that if $d_{J-a\
M'}^{J}(-\frac{\pi}{2})$ is not equal to zero, then $d_{J-a\
-M'}^{J}(-\frac{\pi}{2})$ neither and vice versa. So there is an
epsilon neighborhood of $\theta=-\frac{\pi}{2}$, where $d_{J-a\
M'}^{J}(\theta)$ and $d_{J-a\ -M'}^{J}(\theta)$ are not equal to
zero, then \begin{align*}
\left.\frac{\partial}{\partial\theta}\left(\left|d_{J-a\
M'}^{J}(\theta)\right|+\left|d_{J-a\
-M'}^{J}(\theta)\right|\right)\right|_{\theta=-\frac{\pi}{2}}=0.\end{align*}

\end{enumerate}
It is easy to see\begin{align*}
\frac{\partial}{\partial\theta}\left.\left(\sum_{M'=-J}^{M'=J}\left|d_{MM'}^{J}(\theta)\right|\right)\right|_{\theta=-\frac{\pi}{2}}
& =0\end{align*} $\left.\frac{\partial}{\partial\theta}\left|d_{J-a\
0}^{J}(\theta)\right|\right|_{\theta=-\frac{\pi}{2}}$ only exist
when $2J$ is even. Finally we proof that
$\sum_{M'=-J}^{M'=J}\left|d_{MM'}^{J}(\theta)\right|$reaches its
extreme value when $\theta=-\frac{\pi}{2}$.

\subsection{$\theta=\pi$}

If we substitute $\theta=\pi$ to Eq. (\ref{eq:d_general}), many
items in the right side will be vanished except the one  satisfies
$J+a+M'-2\chi$. Therefore, we can derive

\begin{align*}
d_{J-a\ M'}^{J}(\pi) & =(-1)^{J-a-M'+2\chi}AB(\chi_{\pi})\\
 & =(-1)^{2J}AB(\chi_{\pi}),\end{align*}
where $\chi_{\pi}=\frac{1}{2}(J+a+M')$. With the same consideration,
we substitute $\theta=\pi$ to Eq.
(\ref{eq:partial_d}):\begin{align*}
\left.\frac{\partial}{\partial\theta}d_{J-a\
M'}^{J}(\theta)\right|_{\theta=\pi} &
=\frac{1}{2}(-1)^{2J}AB(\chi_{\pi'}),\end{align*} where
$\chi_{\pi'}=\frac{1}{2}(J+a+M'-1)=\chi_{\pi}-1/2$. As we mentioned
before, $\chi$ is an arbitrary integer. It means $\chi_{\pi}$ or
$\chi_{\pi'}$ can not be established.
\begin{enumerate}
\item If we assume $\chi_{\pi}$ is not satisfied, then $d_{J-a\ M'}^{J}(\pi)=0$
and $|d_{J-a\ M'}^{J}(\pi)|=0$ indicate that $\theta=\pi$ will make
$d_{J-a\ M'}^{J}(\theta)$ reach its extremum value.
\item If we assume $\chi_{\pi'}$ is not satisfied, then $\left.\frac{\partial}{\partial\theta}d_{J-a\ M'}^{J}\right|_{\theta=\pi}=0$.
If $d_{J-a\ M'}^{J}(\pi)=0$, then $\theta=\pi$ is the extremum
point. If $d_{J-a\ M'}^{J}(\pi)\neq0$, then there is an epsilon
neighborhood of $\theta=\pi$, where $d_{J-a\ M'}^{J}(\theta)\neq0$,
so $\left.\frac{\partial}{\partial\theta}\left|d_{J-a\
M'}^{J}(\theta)\right|\right|_{\theta=\pi}=\left.\frac{\partial}{\partial\theta}d_{J-a\
M'}^{J}\right|_{\theta=\pi}=0.$ $\theta=\pi$ is still the extremum
point.
\end{enumerate}
If every $\left|d_{J-a\ M'}^{J}(\theta)\right|$ can achieve its
extremum value when $\theta=\pi$, then
$\sum_{M'=-J}^{M'=J}\left|d_{MM'}^{J}(\theta)\right|$ can also
achieve its extremum value.

\subsection{$\theta=-\pi$}

Substituting $\theta=-\pi$ to Eq. (\ref{eq:d_general})

\begin{align*}
d_{J-a\ M'}^{J}(-\pi) & =AB(\chi_{\pi})\\
 & =AB(\chi_{\pi}),\end{align*}
where $\chi_{\pi}=\frac{1}{2}(J+a+M')$. Substituting $\theta=-\pi$
to Eq. (\ref{eq:partial_d}):\begin{align*}
\left.\frac{\partial}{\partial\theta}d_{J-a\
M'}^{J}(\theta)\right|_{\theta=-\pi} &
=\frac{1}{2}AB(\chi_{\pi'}),\end{align*} Where
$\chi_{\pi'}=\chi_{\pi}-1/2$. This is the same situation as the case
$\theta=\pi$. Through the same discussion,
$\sum_{M'=-J}^{M'=J}\left|d_{MM'}^{J}(\theta)\right|$ can also
attain its extremum value when $\theta=-\pi$.

\subsection{$\theta=0$}

Substituting $\theta=0$ to Eq. (\ref{eq:d_general})

\begin{align*}
d_{J-a\ M'}^{J}(0) & =AB(\chi_{0})\\
 & =AB(\chi_{0}),\end{align*}
where $\chi_{0}=-\frac{1}{2}(J-a-M')$. Substituting $\theta=0$ to
Eq. (\ref{eq:partial_d}):\begin{align*}
\left.\frac{\partial}{\partial\theta}d_{J-a\
M'}^{J}(\theta)\right|_{\theta=0} &
=-\frac{1}{2}AB(\chi_{0'}),\end{align*} Where
$\chi_{0'}=-\frac{1}{2}(J-a-M'-1)=\chi_{0}+1/2$. The analyze will be
the same as $\theta=\pm\pi$. It is easy to prove that
$\sum_{M'=-J}^{M'=J}\left|d_{MM'}^{J}(\theta)\right|$ can also
achieve its extremum value when $\theta=0$.


\end{document}